\documentclass[prx,twocolumn,amsmath,amssymb,usenames,dvipsnames,svgnames]{revtex4}

\usepackage{graphicx}
\usepackage{dcolumn}
\usepackage{bm}
\usepackage{mathrsfs}
\usepackage{subfigure}
\usepackage{natbib}
\usepackage{amsmath}
\usepackage{amssymb}
\usepackage{Jonasmacros}
\usepackage{txfonts}
\usepackage{ifsym}
\usepackage{float}
\usepackage{braket}
\usepackage{epstopdf} 
\usepackage[colorlinks,citecolor=red,linkcolor=Fuchsia]{hyperref}
\usepackage{xcolor}
\usepackage{tikz}

\newcommand{\tobedeleted}[1]{\textcolor{green}{#1}}
\renewcommand{\tobedeleted}[1]{\relax}

\newcommand{\Hbm}{\mathcal{H}}
\newcommand{\beq}{\begin{equation}}
\newcommand{\eeq}{\end{equation}}
\newcommand{\beqa}{\begin{align}}
\newcommand{\eeqa}{\end{align}}

\begin{document} 
\date{\today} 

\title{Nonequilibrium Induced Molecular Quantum Coherence Drives Magnetic Ordering in Coordination Compounds}
\author{Jhoan Alexis Fernandez Sanchez}
\author{Luis Alejandro Sierra Ossa}
\affiliation{Departamento de Fisica, Universidad del Valle, A.A. 25360, Cali, Colombia.}
\author{Henning Hammar}
\author{Jonas Fransson}
\affiliation{Department of Physics and Astronomy, Uppsala University, Box 516, SE-751 21 \ UPPSALA, Sweden}
\author{Juan David Vasquez Jaramillo}
\affiliation{Departamento de Fisica, Universidad del Valle, A.A. 25360, Cali, Colombia.}
\affiliation{Grupo de Optica Moderna - Departamento de Fisica y Geologia, Universidad de Pamplona}
\affiliation{Instituto de Fisica, Universidad de Antioquia. Medellin Colombia}
\email{juan.vasquez@unipamplona.edu.co}

\begin{abstract}
Novel understanding of the recent nanomagnet tailoring experiments and the possibility to further unveil the mechanisms by which the magnetic interactions arise in an atom by atom fashion covers importance as the demand for spin qubit and quantum state detection architectures increases. Here, we address the spin states of a molecular trimer comprising three localized spin moments embedded in a metallic tunnel junction and show that the pair spin interactions can be engineered through the electronic structure of the molecular trimer. We show that bias and gate voltages induce either a completely ferromagnetic state of the localized moments or a spin frustrated state with different stabilities, and that switching between these states is possible on demand by electrical control. The role of quantum coherence in the molecular trimer is discussed with regards to the spin ordering as well as the interplay among electronic interference and induced dephasing by the metallic leads. This work sets foundations for more robust all electrically controlled spin architectures usable in quantum engineering systems and serves as a test bench for exploring unresolved questions in magnetic ordering and symmetry.   
\end{abstract} 
\pacs{}

\maketitle
\section{Introduction}
Engineering magnetism at the atomic scale has emerged as one of the most vibrant, interesting and challenging areas of study in the field of condensed matter in the last decade. Fuelled by novel experimental techniques developed in the context of scanning tunneling microscopy (STM), scientists have been empowered to manipulate, control and probe matter at atomic scale resolution \cite{Chen2007,Eigler1990}. The manipulation of atomic spins has yet captured much of the focus and efforts in experimental nanotechnology \cite{Loth2012,Khajetoorians2012,Heinrich2013}. Driven by the need to produce novel quantum states of matter optimal for the use in quantum information and quantum engineering applications as well as for emergent spintronic technology \cite{Heinrich2015,Bogani2009,Urdampilleta2011}, it is desired to tune spin-spin interactions to either ferromagnetic or to antiferromagnetic alignment  \cite{Khajetoorians2012,Wagner2013} with stable single spin anisotropy.

The magnetic interactions, both in sign and magnitude, which respectively provide ferro- or antiferromagnetic couplings between spins, as well as magnetic anisotropies, can be adjusted through indirect spin interactions by properly orienting and separating the magnetic ions in the molecule being adsorbed \cite{Khajetoorians2012}. On the other hand, the indirect interactions are completely determined by the electronic structure of the substrate, what suggests that engineering the density of states of the host will influence the sign and magnitude of the effective exchange coupling among the spins grafted onto it through the Kondo interaction \cite{Khajetoorians2012,Wagner2013,Khajetoorians2011,Simon2011}. 

By invoking the results in \cite{Koole2015,Xia2015,Tsuji2016,Fracasso2011,Pedersen2015}, the transport properties of molecular junctions, particularly the (differential) conductance, can be hindered or enhanced depending on the nature and origin of quantum interference present within the molecular structure of interest. A manifestation of this has been shown to be intimately related with the conjugation order of the molecule, whether broken, cross or linear \cite{Bergfield2009, Valkenier2014,Guedon2012}, and its bonding position with the electrodes or with the linking group, weather para, meta or ortho \cite{Pedersen2015}.

Theoretical predictions have confirmed the observed low conductance in junctions under the influence of destructive quantum interference or even in the presence of quantum decoherence \cite{Saygun2016,Jaramillo2017}, which is a condition that as well arises in magnetic tunneling junctions of a dimer of spins resembling the experiment reported in \cite{Urdampilleta2011}, when both units are anti-ferromagnetically coupled. Following the same logic, it should in principle be possible to engineer spin-spin interactions in magnetic tunneling junctions by controlling the degree of quantum interference and the emergence of electronic decoherence allowed by the density of electron states, both in sign and magnitude. Here, the fully anti-ferromagnetic interaction is expected to localize the electron states in each energy level available for occupation whereas the fully ferromagnetic interaction is expected to delocalized the electron wave in the molecule, hence, favoring quantum coherence of different nature \cite{Saygun2016}.

In this context, according to the work of \cite{Sharples2014,Wu2017,Grindell2016,Khajetoorians2012} where frustrated spin geometries have been engineered, completely ferromagnetic ordered, have been tailored, investigations on the effect of electronic quantum coherence and the emergence of decoherence is of great relevance and impact for the studies of open questions in atomic magnetism and in designing building blocks for quantum engineering systems. These issues pose giant challenges for experiments and theory.

The above arguments converge in the grounds of tailoring effective spin ordering by the coordinated action of gate and bias potentials, giving rise of what we call the magnetic $V_\text{SD}$--$V_\text{G}$ diagram. Subsequently these interactions can be probed using all-electrical measurements as predicted in \cite{Fransson2014}, a completely unexplored issue both from the perspective of non-equilibrium nanophysics and from the point of view of nanotechnology and manipulation at the atomic scale.

\section{Model}
\begin{figure}[t]
\begin{center}
\includegraphics[width=\columnwidth]{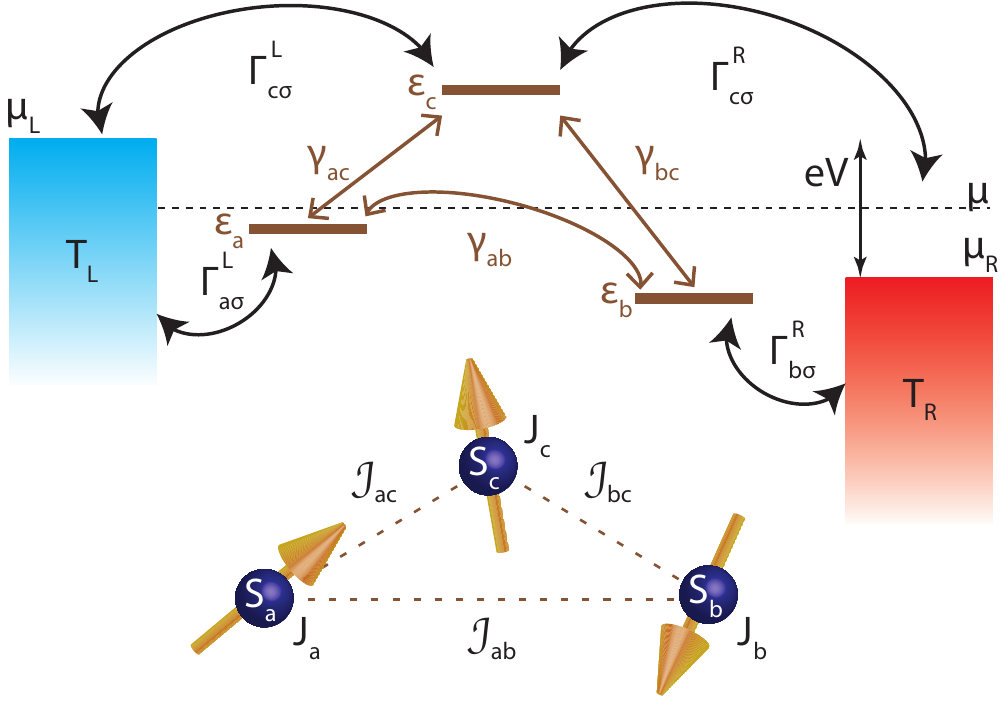}
\end{center}
\caption{Molecular magnet composed by a trimer of spins: This illustration describes a molecular trimer of electronic levels $\epsilon_{m\sigma}$, for $m=A,B,C$, each of them coupled to a local spin moment $\bm{S}_{m}$ through the Kondo interaction of strength $J_{m}$. Here, the Kondo interaction between the electronic level and the localized spin is not shown in arrow form due to lack of space and to keep clarity on where the interactions are and how they are labeled. The interaction $\bm{\gamma}_{mn}$, for $m\neq n$, $m,n=a,b,c$, represents the hopping amplitude for an electron in level $\epsilon_{m\sigma}$ to undergo a transition to level $\epsilon_{n\sigma}$. This hopping amplitude is represented in matrix form by a symmetric tensor denoted as $\left[\bm{\gamma}\right]$.  The couplings to the leads, both left ($\alpha$) and right ($\beta$), are represented by the matrix element $\bm{\Gamma}^{(\alpha,\beta)}_{mn\sigma\sigma}$, and for the diagonal matrix elements ($m=n$), these couplings represent the tunneling amplitude for an electron in the left or right lead to transit into level $\epsilon_{m\sigma}$. The off-diagonal matrix elements not shown in the diagram, represent co-tunneling processes through different electronic levels}
\label{three_spins}
\end{figure}

Here, in the present work, we consider a trimer of magnetic ions immersed in an organic molecule that can be a phthalocyanine, a metal hydride such as M-porphyrin, where  M denotes a transition metal element, or an organometal \cite{Jaramillo2017,Saygun2016}, exhibiting the possibility of quantum interference whether destructive or constructive. The magnetic ions, which display a localized magnetic moment, do not interact directly, but only through the electron interactions in the host molecule, therefore, the spin-spin interactions are tuned via the molecular electronic structure. The magnetic ion-molecule system is adsorbed onto a metallic substrate, and then probed by a metallic STM tip. As such the Hamiltonian of the system STM--magnetic ion--molecule--substrate is given by:
\begin{subequations}
\label{ham}
\begin{align}
\mathcal{H}&=\mathcal{H}_{tip}+\mathcal{H}_{ion}+\mathcal{H}_{mol}+\mathcal{H}_{subst}+\mathcal{H}_{hyb},
\label{ham1}
\\
\Hbm_{tip}&=\sum_{\bm{k}\sigma}\epsilon_{\bm{k}\sigma}(t)c_{\bm{k}\sigma}^{\dagger}(t)c_{\bm{k}\sigma}(t),
\label{Dimer_Ham1b}
\\
\Hbm_{subst}&=\sum_{\bm{q}\sigma}\epsilon_{\bm{q}\sigma}(t)c_{\bm{q}\sigma}^{\dagger}(t)c_{\bm{q}\sigma}(t),
\label{Dimer_Ham1a}
\\
\Hbm_{mol}&=\sum_{m\sigma}\epsilon_{m\sigma}d_{m\sigma}^{\dagger}(t)d_{m\sigma}(t)
+\sum_{mn\sigma}\gamma_{mn}d^{\dagger}_{m\sigma}(t)d_{n\sigma}(t)
\label{Dimer_Ham1c}
\\
\Hbm_{sd}&=\sum_{m}J_{m}\bm{s}_{m}(t)\cdot\bm{S}_{m}(t),
\label{Dimer_Ham2}
\\
\Hbm_{tip-mol}&=\sum_{m\bm{k}\sigma}v_{m\bm{k}\sigma}(t)c_{\bm{k}\sigma}^{\dagger}(t)d_{m\sigma}(t)+H.c.
\label{Dimer_Ham3a}
\\
\Hbm_{subst-mol}&=\sum_{m\bm{q}\sigma}v_{m\bm{q}\sigma}(t)c_{\bm{q}\sigma}^{\dagger}(t)d_{m\sigma}(t)+H.c.
\label{Dimer_Ham3b}
\end{align}
\end{subequations}
The system described through the Hamiltonian given in Eq. (\ref{ham}), is illustrated in Fig. \ref{three_spins}.

In the model presented by Eq. (\ref{ham}), $\Hbm_{tip}$ represents the Hamiltonian of the metallic tip, $\Hbm_{sust}$ the Hamiltonian for the metallic substrate, $\Hbm_{mol}$ is the molecular Hamiltonian, the local interaction between the magnetic ion and the conduction electrons is given by $\Hbm_{sd}$ and the Tip-Molecule and Substrate-Molecule hybridization are given by Hamiltonians $\Hbm_{tip-mol}$ and $\Hbm_{subst-mol}$. The parameters defined in the model Hamiltonian are defined as follows: $\epsilon_{\bm{k}\sigma}(t)$ and $\epsilon_{\bm{q}\sigma}(t)$ are the energy bands of the metallic tip and the metallic substrate respectively, $\epsilon_{m\sigma}$ is the energy of the $m-th$ molecular orbital, $J_{m}$ is the Kondo coupling between the $m-th$ spin moment of the electron with energy $\epsilon_{m\sigma}$ and the $m-th$ localized spin moment denoted as $\bm{S}_{m}(t)$, and, $v_{m\bm{k}\sigma}(t)$ and $v_{m\bm{q}\sigma}(t)$ are the overlap integrals between wave functions from the tip and the molecule, and the substrate and the molecule respectively. 

The Green's function for the model represented by Eq. (\ref{ham}), can be obtained from the inverse of the retarded Green's function given by expression \ref{spin3}. By inverting equation \ref{inv}, the retarded Green's function is evaluated, and the lesser and greater Green's functions are obtained from
\begin{align}
G^{</>}_{mn\sigma\sigma'}(\omega)&=\frac{(\pm i)}{\hbar}\sum_{\chi\mu\nu}f^{(\chi)}_{\sigma}(\pm\epsilon)G_{m\mu\sigma\sigma'}(\omega)\Gamma^{(\chi)}_{\mu\nu\sigma}G^{A}_{\nu n\sigma\sigma'}(\omega),
\label{gless}
\end{align}
where $\epsilon$ and $\omega$ are related through $\epsilon=\hbar\omega$ and $\chi$ indexes the lead, whether $\alpha$ for the left lead, or $\beta$ for the right lead. In Eq. (\ref{gless}), the matrix $\bm{\Gamma}^{(\chi)}_{\sigma}$ is proportional to the imaginary part of the retarded self-energy $\bm{\Sigma}^{(\chi)R}_{\sigma}$ which can be paramterized by (see appendix \ref{green_func})
\begin{equation} 
\bm{\Sigma}^{(\chi)R}_{\sigma}=\bm{\Lambda}_{\sigma}^{(\chi)}-\frac{i}{2}\bm{\Gamma}^{(\chi)}_{\sigma}.
\label{self_sigma}
\end{equation}
Here, $\bm{\Lambda}_{\sigma}^{(\chi)}$ is related to the Lamb-Shift \cite{Esposito2015a,Esposito2015b} and $\bm{\Gamma}_{\sigma}^{(\chi)}$ represent the coupling between the leads and the molecular energy levels. The diagonal matrix elements of the latter stands for the level broadening, and the off-diagonal matrix elements are related to the dephasing among levels coupled to the reservoir \cite{Jauho1994,Bedkihal2012,Matityahu2013,Tu2014}. To determine the form of matrix element $\bm{\Gamma}^{(\chi)}_{mn\sigma}$, we consider the definition from \cite{Jauho1994} of the retarded self-energy matrix element:
\begin{align} 
\bm{\Sigma}^{(\chi)R}_{mn\sigma}&=\sum_{\bm{k}}\frac{v_{m\bm{k}\sigma}v_{n\bm{k}\sigma}^{*}}{\epsilon-\epsilon_{\bm{k}\sigma}+i\delta},
\nonumber
\\
&={\cal P}\sum_{\bm{k}}\frac{v_{m\bm{k}\sigma}v_{n\bm{k}\sigma}^{*}}{\epsilon-\epsilon_{\bm{k}\sigma}}-i\pi\sum_{\bm{k}}\delta(\epsilon-\epsilon_{\bm{k}\sigma})v_{m\bm{k}\sigma}v_{n\bm{k}\sigma}^{*}.
\label{self_sigma1}
\end{align}
In the above expression, the couplings $v_{m\bm{k}\sigma}$ and $v_{n\bm{k}\sigma}^{*}$ appearing in the model Hamiltonians given by Eqs. (\ref{Dimer_Ham3a}) and (\ref{Dimer_Ham3b}), are complex in nature, and therefore, we can express both of them as an amplitude times a phase factor of the form $v_{m\bm{k}\sigma}=\left|v_{m\bm{k}\sigma}\right|e^{-i\phi_{m}}$ and $v_{m\bm{k}\sigma}^{*}=\left|v_{m\bm{k}\sigma}\right|e^{i\phi_{m}}$, to transform Eq. (\ref{self_sigma1}) into
\begin{align} 
\bm{\Sigma}^{(\chi)R}_{mn\sigma}&=\sum_{\bm{k}}\frac{\left|v_{m\bm{k}\sigma}\right|\left|v_{n\bm{k}\sigma}\right|e^{-i(\phi_{m}-\phi_{n})}}{\epsilon-\epsilon_{\bm{k}\sigma}+i\delta}.
\label{self_sigma2}
\end{align}
From Eq. (\ref{self_sigma2}), we thus, define the elements $\Gamma^{(\chi)}_{mn\sigma}$ according to
\begin{align}
\Gamma^{(\chi)}_{mn\sigma}&=2\pi\sum_{\bm{k}}\delta(\epsilon-\epsilon_{\bm{k}\sigma})\left|v_{m\bm{k}\sigma}\right|\left|v_{n\bm{k}\sigma}\right|e^{-i(\phi_{m}-\phi_{n})},
\label{self_sigma4}
\end{align}
where phases $\phi_{m}$ and $\phi_{n}$ determine the strength of the dephasing between levels $m$ and $n$. Moreover, it becomes crucial to determine the density of electron states $\bm{\rho}_{c}(\omega)$ in the quest for an understanding on what is the interplay between electronic structure and magnetism in the sample. The former can be obtained from the equation
\begin{equation}
\bm{\rho}_{c}(\omega)=\frac{i}{2\pi}\sum_{m\sigma}\sigma_{\sigma\sigma}^{(0)}\left(\bm{G}^{>}_{mm\sigma}(\omega)-\bm{G}^{<}_{mm\sigma}(\omega)\right),
\label{dos1}
\end{equation}
and similarly, the spin density of states
\begin{equation}
\bm{\rho}_{s}(\omega)=\frac{i}{2\pi}\sum_{m\sigma}\sigma_{\sigma\sigma}^{(z)}\left(\bm{G}^{>}_{mm\sigma\sigma}(\omega)-\bm{G}^{<}_{mm\sigma\sigma}(\omega)\right).
\label{dos2}
\end{equation}

In addition to the latter, it is often useful to think about magnetic ordering and change in magnetic configuration in terms of magnetic entropy and the corresponding energy flow, which holds intimacy with the symmetry of the system through which flows. By invoking the von Neuman entropy $\bm{\mathcal{S}}_{\sigma}$ \cite{Breuer2007} given by
\begin{equation}
\bm{\mathcal{S}}_{\sigma}=-\bm{\rho}_{\sigma}\ln \bm{\rho}_{\sigma},
\label{entropy}
\end{equation}
where $\bm{\rho}_{\sigma}$ denotes the density matrix of the system, we can write, a general expression in terms of contour Green's functions for the entropy associated with spin degrees of freedom can be written
\begin{equation}
\bm{\mathcal{S}}_{\sigma}=-i\hbar\int \tr\left[\bm{G}^{<}_{\sigma\sigma}(\omega)\ln(-i\hbar)\bm{G}^{<}_{\sigma\sigma}(\omega)\right]\frac{d\omega}{2\pi}.
\label{entropy1}
\end{equation}
The above formulation for the spin entropy derives from the von Neumann expression obtained from Eq. (\ref{entropy}), where the density matrix has been defined with the aid of the contour ordered Green's function $\bm{G}^{<}_{\sigma\sigma}(\omega)$ in matrix form
\begin{equation}
\bm{\rho}_{\sigma}=\left\langle \left(\begin{array}{ccc}  d^{\dagger}_{a\sigma}&d^{\dagger}_{b\sigma}&d^{\dagger}_{c\sigma}\end{array}\right)\left(\begin{array}{c} d_{a\sigma}\\ d_{b\sigma}\\ d_{c\sigma}\end{array}\right)\right\rangle=
-i\hbar \bm{G}^{<}_{\sigma\sigma}(\omega).
\label{density}
\end{equation}

Here we approach the question of the emergence of quantum interference and its relationship to the spin configuration of the molecule of interest. A useful theoretical tool available from the Landauer formalism \cite{Jauho1994} to determine the degree of quantum interference in a molecular conductor is the Transmission probability given by
\begin{equation}
\mathcal{T}_{\alpha\beta}(\epsilon)=\tr G^{R}(\omega)\Gamma^{(\alpha)}G^{A}(\omega)\Gamma^{(\beta)}.
\label{transx}
\end{equation}
From the same viewpoint, other theoretical tools such as particle and energy currents have lead to useful predictions in systems relevant for the discussion of the present paper \cite{Wagner2013,Saygun2016,Jaramillo2017}, and therefore we have considered them into the investigation here reported. These transport quantities are written according to
\begin{subequations}
\begin{align}
\mathcal{I}_{N}^{(\chi)}=&
	-\frac{1}{h}\sum_{\sigma}\int\Bigl(f_{\chi\sigma}(\dote{})-f_{\bar\chi\sigma}(\dote{})\Bigr){\cal T}_{\chi\bar\chi}(\dote{})d\dote{}
	,
\\
\mathcal{I}_{E}^{(\chi)}=&
	-\frac{1}{h}\sum_{\sigma}\int \dote{}\Bigl(f_{\chi\sigma}(\dote{})-f_{\bar\chi\sigma}(\dote{})\Bigr){\cal T}_{\chi\bar\chi}(\dote{})d\dote{}
	,
\end{align}
\end{subequations}
where $\bar\chi$ refers to the lead on the opposite side to $\chi$ of the junction.

\vspace{0.4cm}

Along the same line, we study the quantum interference present in the molecule with regards of the trimmer spin structure with an analytic expression for the differential charge conductance $\bm{\sigma}^{\chi} = \partial J_{\chi}^{e} / \partial V$. This analytical observable is found to be given by (see appendix \ref{ConductanceCalculation}):
\begin{align}
    \bm{\sigma}^{\chi}=\mp \, \frac{i\,e^{2}}{2} \int d\epsilon \:  \frac{\beta}{4} \: \cosh^{-2} \left(  \frac{\beta (\epsilon-\mu_{\chi})}{2}  \right)\:Tr \bigg\{ \;  \Gamma^{\chi}  \Big[ \; G^{>}(\epsilon)-G^{<}(\epsilon) \nonumber \\
    + i\,G^{R}(\epsilon)\,\Gamma^{\chi}(\epsilon)\,G^{A}(\epsilon) \Big ]   \bigg\}  \nonumber  \\ 
    \label{FINAL1}
\end{align}
\section{Spin-Spin Effective Interactions and Electronic Quantum Interference}
Effective spin-spin interactions are calculated from the expression derived in ref. \cite{Fransson2014}. Specifically, in this paper we address the effective Heisenberg exchange interactions present in the system described by the model given in Eq. (\ref{ham}) and illustrated in Fig. \ref{three_spins}. These interactions among three localized magnetic moments labeled as $\bm{S}_{a}$, $\bm{S}_{b}$ and $\bm{S}_{c}$, that are coupled via Kondo interactions $J_{a}$, $J_{b}$, and $J_{c}$ with electrons present in three energy levels $\epsilon_{a}$, $\epsilon_{b}$ and $\epsilon_{c}$ respectively, are given by Eq. (\ref{exchange}). Here, we pay special attention to two cases of great interest, which are the ferromagnetic state where all effective interactions $\bm{\mathcal{J}}_{mn}$ are negative, and the antiferromagnetic state where these interactions are now all positive. The stability and control of these ordering, as well as its manipulation and detection can be done by several means, one of them being the all-electrical control as demonstrated for Manganese based metal hydrides \cite{Osorio2010}, among other experimental realizations \cite{Vincent2012,Aradhya2013,Urdampilleta2011,Krainov2017,Mannini2014}. All-electrical control has also been shown to allow for atom by atom tailoring of nanomagnets \cite{Khajetoorians2012} thorugh the RKKY interaction including spin-frustrated networks. Moreover, all electrical control has provided a means of control for the singlet-triplet switching in a dimer of Cobalt atoms, and these states were detected by measurements of charge current flowing through the host molecule. In single magnetic unit molecules, important properties have been also engineered with the aid of a bias voltage. Another degree of control possible is the control through gate fields, demonstrated in the Anthraquinone transistor, in which case the destructive quantum interference exhibited by this type of molecule was lifted by the action of this gate \cite{Koole2015}. The latter suggests that a combined bias voltage - gate field control scheme can  provide the possibility to switch between spin states determined by the magnetic ordering present in the molecule, and the nature and strength of quantum interference have a strong chance to also play a role in this switching dynamics. The system we address here, combines these means of control, that is, the electric control provided by a bias voltage and control driven by the gate field, as well as the variation of the degree of quantum coherence present in the electronic part of the molecule. The latter is achieved by means of modulating the model given by Eq. (\ref{ham}) with associated Green's function given by Eq. (\ref{inv}), in terms of the parameter $\gamma_{ab}$, which varies from $0$ to $6.0$ meV, where $\gamma_{ab}=0$ meV resembles a molecular structure with no possibility to exhibit quantum interference, and as $\gamma_{ab}$ increases, the degree of electronic coherence in the system increases. This way of controlling quantum interference in the multiple sites model in the electronic $\Lambda$--$\Delta$ system was studied in ref. \cite{Guedon2012}. Here, we extend the investigation by considering a molecule with magnetic units, the question of atom by atom engineered exchange interaction and the interplay among the degree of electronic quantum interference and other means of control, namely a bias voltage and a gate field, in the switching dynamics of magnetic ordering in the molecule of interest. (See Fig. \ref{three_spins})

\begin{figure}[t]
\begin{center}
\includegraphics[width=\columnwidth]{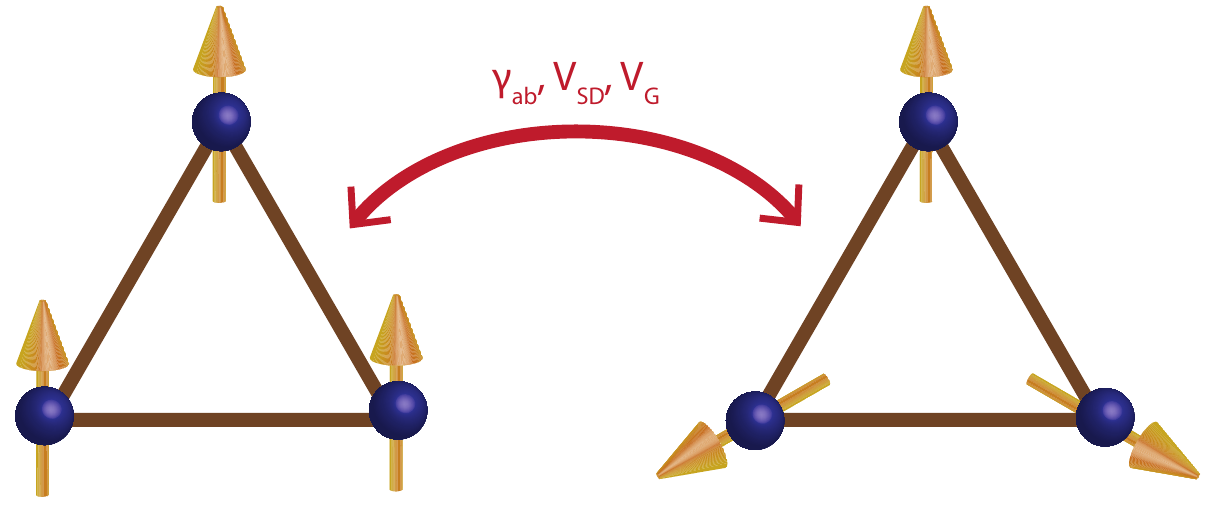}
\end{center}
\caption{Delocalization induced switching dynamics of the spin ordering in the molecular trimer shown in Fig. \ref{three_spins}. Once the spin-frustrated state is formed in the molecule, the ferromagnetic alignment in the system can be obtained by tuning the intermolecular coupling $\gamma_{ab}$, the source-drain voltage $V_\text{SD}$, or the gate voltage $V_\text{G}$.}
\label{schemes1}
\end{figure}

\begin{figure}[b]
	\centering
		\includegraphics[width=\columnwidth]{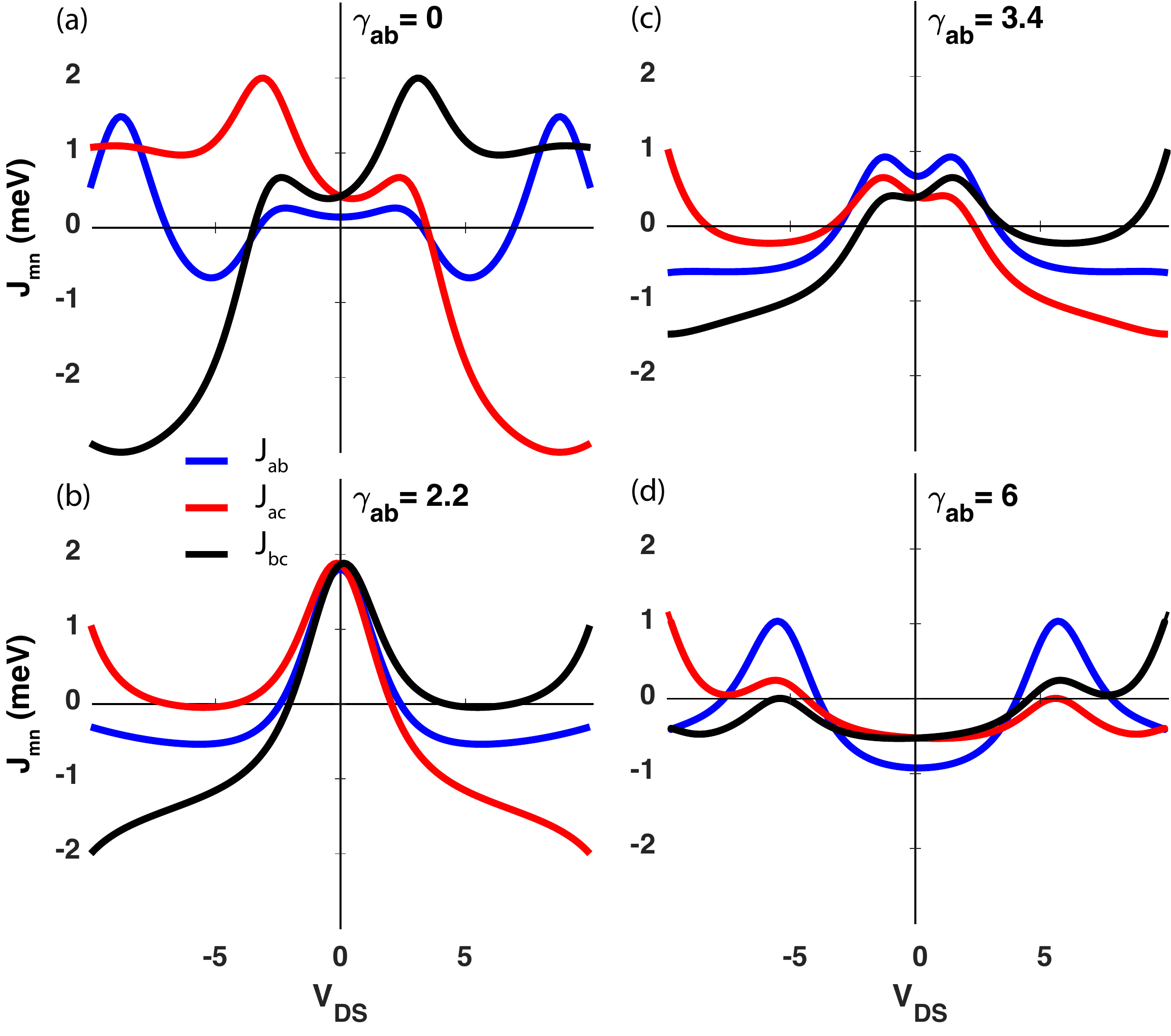}
	\caption{Effective exchange interactions $\bm{\mathcal{J}}_{ab}$, $\bm{\mathcal{J}}_{ac}$ and $\bm{\mathcal{J}}_{bc}$ among spins $\bm{S}_{a}$, $\bm{S}_{b}$ and $\bm{S}_{c}$. The upper left panel presents the effective exchange interactions $\bm{\mathcal{J}}_{mn}$ for $m,n=a,b,c$ for the case in which 
	$\gamma_{ab}=0.0$ mV, exhibiting a weak spin ordering arround zero-bias, with $V_\text{G}=-4.0$ mV. The strength of this ordering increases as $\gamma_{ab}$ varies from 2.2, 3.4, and 6.0 meV, switching from a frustrated configuration to a ferromagnetic state.}
	\label{exchange1}
\end{figure}

\begin{figure}[b]
	\centering
		\includegraphics[width=\columnwidth]{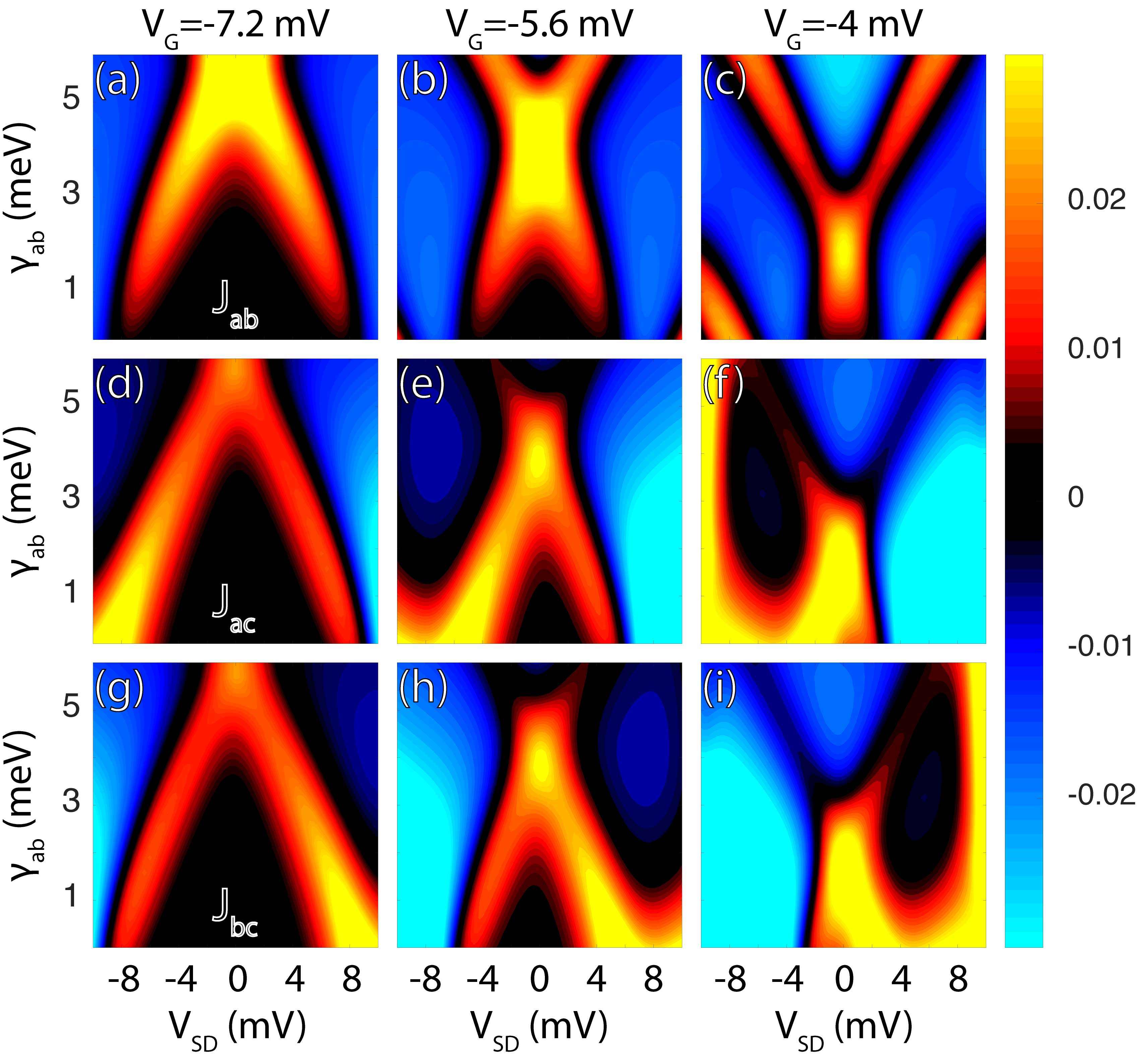}
	\caption{Magnetic diagram $V_\text{SD}$--$\gamma_{ab}$. It shows the regions where magnetic frustration in the spin system emerges as well as where the ferromagnetic order arises. From the magnetic diagram shown in Fig. \ref{frust6}, it can be inferred that for $V_\text{G}=-4.0$ mV, the Neel state of spin frustration emerges through the systematic variations of $\gamma_{ab}$. For large values of $\gamma_{ab}$ it can be seen that the ferromagnetic state is obtained.}
	\label{gamma_switch}
\end{figure}
\section{Spin Expectation Values in the presence of a Zeeman magnetic field}
With the aim of visualising phase transitions in magnetic ordering of the trimmer, we study the effect of a symmetry breaking phenomenom induced by a Zeeman term in the effective spin-spin Heisenberg interaction Hamiltonian \eqref{eff} by means of the spin expectation values given by \eqref{B14}. The new spin Hamiltonian is:
\begin{equation}
    \mathcal{H}_{spin} = \sum_{m,n} \mathcal{J}_{mn}\, \bm{S}_{m} \cdot \bm{S}_{n} - \Delta \sum_{k} \bm{S}_{k} \cdot \bm{B}_{k}
\end{equation}, where $\Delta$ is a constant equal to one in atomic units, and $\bm{B}_{k}$ are staggered fields for each localized spin.
The isotropic nature of the Heisenberg effective interaction permits to project each localized spin along the z-direction in particular, without loose of generality, in order to characterize the phase transitions.

\section{Results}
The voltage, electric field and delocalization induced switching dynamics of the spin ordering in the molecule of interest in the present paper is demonstrated here, by predicting the variation of the effective exchange interactions among the magnetic units present in the molecules, namely $\bm{\mathcal{J}}_{AB}$, $\bm{\mathcal{J}}_{AC}$ and $\bm{\mathcal{J}}_{BC}$. This variation is shown to be induced by the modulation of three different conditions driving the molecular degrees of freedom. First, parameter $\gamma_{ab}$ is varied in the range $0\leq\gamma_{ab}\leq6$ meV. The variation of this parameter, induces a change in spin ordering in the molecule from nearly eight-fold-degeneracy and spin-frustrated state to all ferromagnetic ordering as depicted in the scheme shown in Fig. \ref{schemes1}, and as predicted through the evaluation of $\bm{\mathcal{J}}_{AB}$, $\bm{\mathcal{J}}_{AC}$ and $\bm{\mathcal{J}}_{BC}$ shown in Fig. \ref{exchange1}, and done by employing Eq. (\ref{exchange}). Moreover, the contour shown in Fig. \ref{gamma_switch} illustrates how the modulation of $\gamma_{ab}$ has an effect on the spin ordering around zero-bias, for $3$ different values for the gate field expressed as an energy $V_\text{G}=-7.2$ mV, -5.6 mV, -4.0 mV. This figure, shows that around zero bias there is a commutation among quantum states with all anti-ferromagnetic spin-spin interactions to those with all ferromagnetic interactions among spins. For $V_\text{G}=-4.0$ mV, the frustrated spin state in the molecules occurs at lower values for the parameter $\gamma_{ab}$, as compare with other values for the gate field. At larger values of $\gamma_{ab}$ for the same gate field $V_\text{G}$, the spins align ferromagnetically. 

\begin{figure}[t]
	\begin{center}
		\includegraphics[width=\columnwidth]{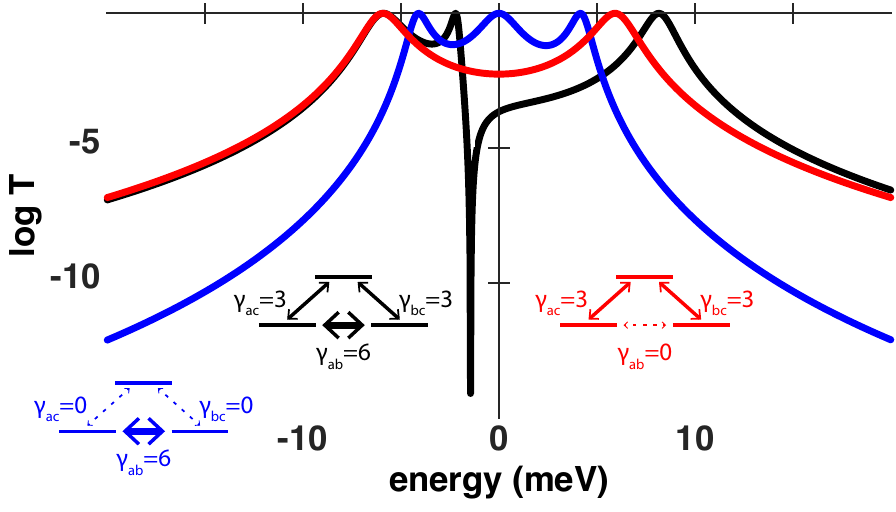}
	\end{center}
	\caption{Transmission probabilities for the model considered in \cite{Guedon2012} for $\gamma_{ac}=\gamma_{bc}=0$ and $\gamma_{ab}=3.0~~meV$ (in blue), $\gamma_{ac}=\gamma_{bc}=3.0$ meV and $\gamma_{ab}=0$ meV (in red) and, $\gamma_{ac}=\gamma_{bc}=3.0$ meV and $\gamma_{ab}=3.0$ meV (in black with yellow stripes). The latter shows a transmission dip around the Fermi level, which is a clear signature of the presence of destructive quantum interference \cite{Guedon2012,Bergfield2009}.}
	\label{model6}
\end{figure}
\begin{figure}[b]
	\centering
		\includegraphics[width=\columnwidth]{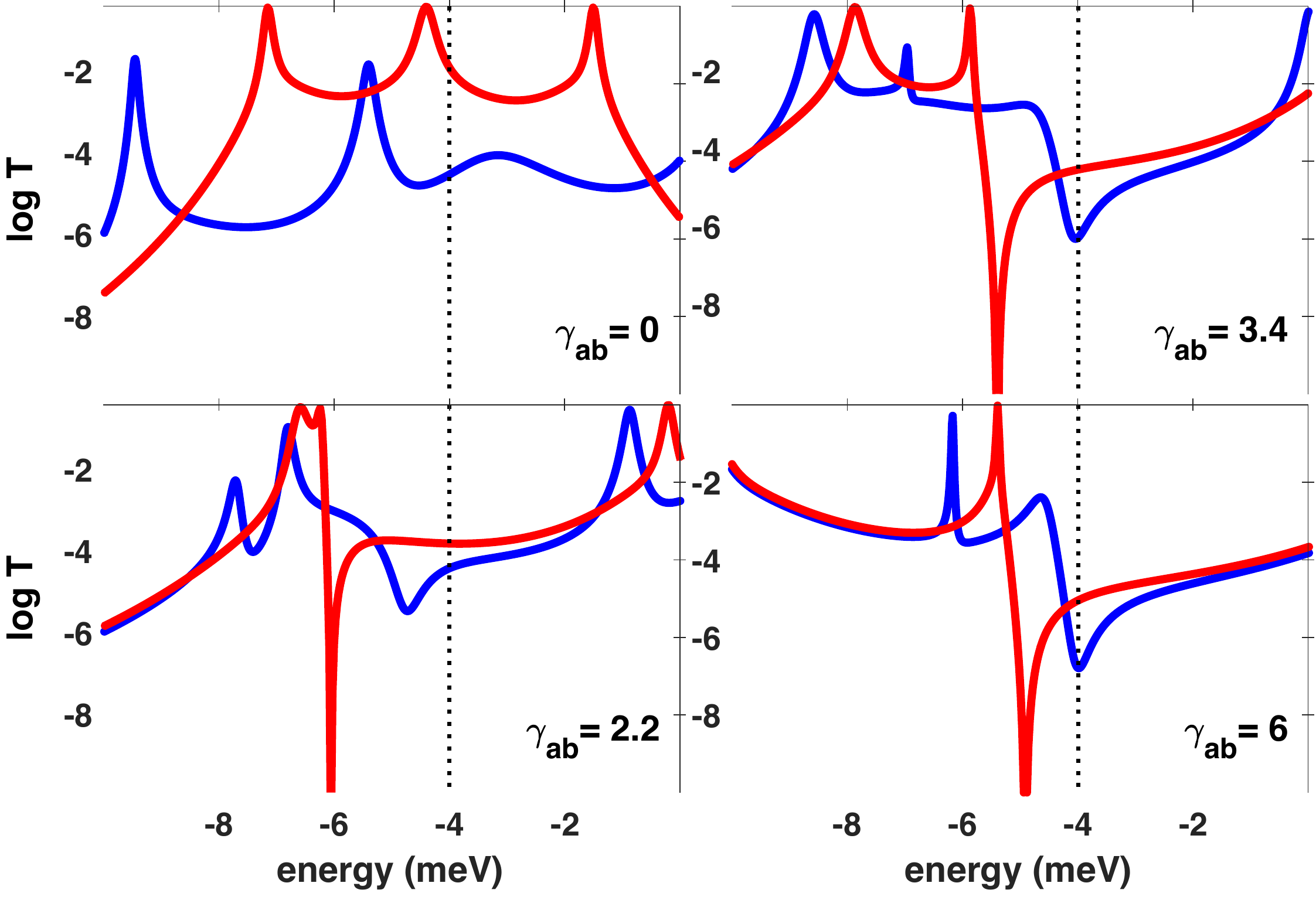}
	\caption{Transmission probabilities for the cases with associated spin ordering shown in Fig. \ref{exchange1}. In red the case where no spin structure is modulating the density of states of the molecule, and in blue, the case in which the spin structure induces variations on the density of states. The dotted black line corresponds to the set value for the gate field, which in this case is $V_\text{G}=-4.0$ mV}
	\label{ttx}
\end{figure}
\begin{figure}[t]
	\centering
		\includegraphics[width=\columnwidth]{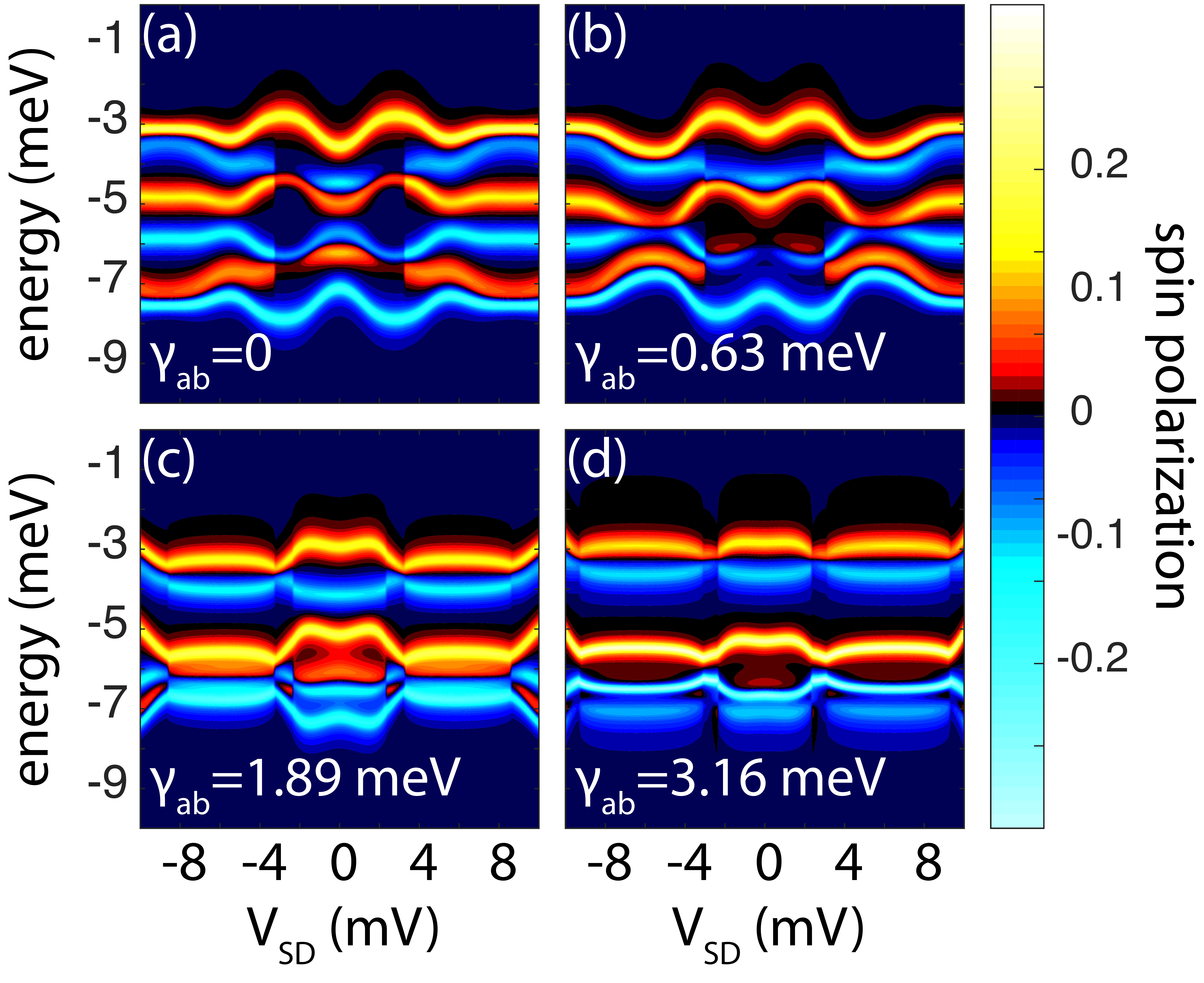}
	\caption{Spin polarization $\bm{\rho}_{s}(\omega)$ of the molecule under study for a gate field $V_\text{G}=-4.0$ mV.}
	\label{spin_dos}
\end{figure}

The phenomenological insights and microscopic mechanisms behind the modulation of the parameter $\gamma_{ab}$, can be understood by considering the phenomenon of quantum interference in molecular junctions as discussed in \cite{Guedon2012}, which investigates quantum coherent phenomena in molecular junction with a similar electronic configuration to the one investigated in this paper, with the differentiating factor of having the couplings $\Gamma^{(\chi)}_{C\sigma}$ for $\chi=\alpha,\beta$ exactly equal to zero.

Under this assumption, the variation of $\gamma_{ab}$ from vanishing behavior to an intermediate value resembles the transition of the molecular structure corresponding to Anthracene-like (linear conjugation) molecular structure, to Anthraquinone-like (Cross-Conjugation) molecular structure where delocalization dominates the electronic processes within the molecule \cite{Guedon2012}. From the electronic transmission probability calculated for the system of interest from Eq. (\ref{transx}), it can be shown that $\gamma_{ab}$ is associated with the ability for the system to exhibit electronic quantum interference as shown in Fig. \ref{model6}, \cite{Valkenier2014}.

By considering the effect of the spin structure in the ability for the system to exhibit quantum interference, mainly of destructive nature, Fig. \ref{ttx} shows that the ordering of the magnetic units in the trimmer has a decisive effect on the strength of the transmission dip, the broadening and the localization in energy, showing that the respective increase in magnetic symmetry is complained with the corresponding exhibition of weak quantum interference of destructive nature, and the corresponding decrease in symmetry, is associated with the lifting of coherence in the molecule.

This particular feature of this system can be understood from considering the spin polarized density of states (see Fig. \ref{spin_dos}) around zero bias and through a bias window in which the system exhibits clear order. In this particular regime, the spin density of states presents abrupt changes when the order is lost, this around the set value for $V_\text{G}$.
\begin{figure}[t]
	\centering
		\includegraphics[width=1.0\columnwidth]{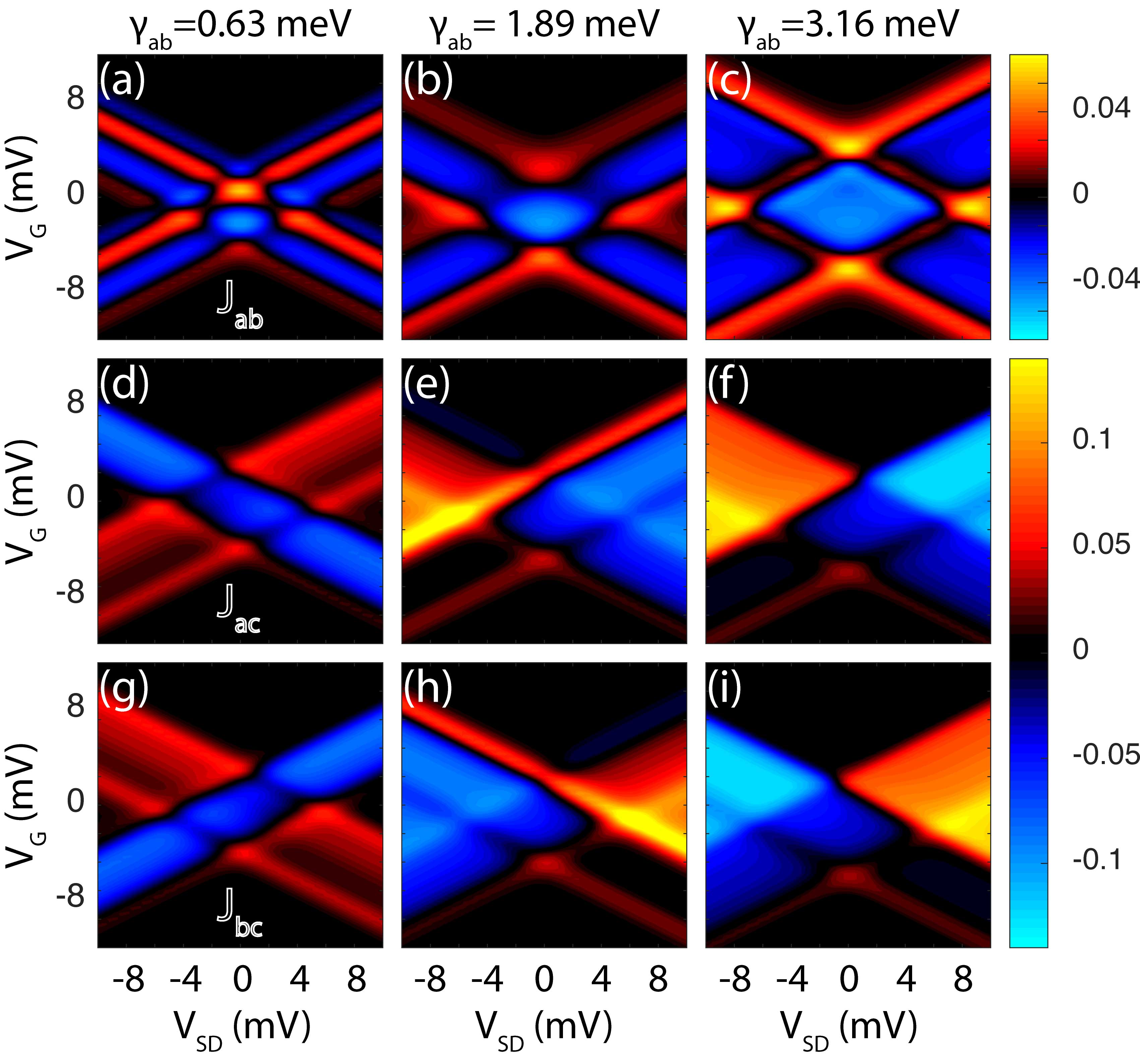}
	\caption{Magnetic $V_\text{SD}$--$V_\text{G}$ diagram. Shows regions where anti-ferromagnetic ordering arises (coincidence in yellow) and other ones where ferromagnetic ordering does it so (coincidence in blue). The panels are arranged in vertical order for different values of the parameter $\gamma_{ab}$.}
	\label{frust6}
\end{figure}

An additional mean for controlling the ordering in the magnetic trimer is the bias voltage $V_\text{SD}$, which induces a non-equilibrium behavior in the system under study. An applied bias will as well induced an anti-ferromagnetic state in the spin trimer by an appropriate driving through the gate field of the molecule, and then, by swapping over a range of bias voltages, this state will be switched to an all ferromagnetic coupling configuration. This scheme of commutation is shown in Fig. \ref{schemes1}. 

\begin{figure}[t]
	\centering
		\includegraphics[width=1.0\columnwidth]{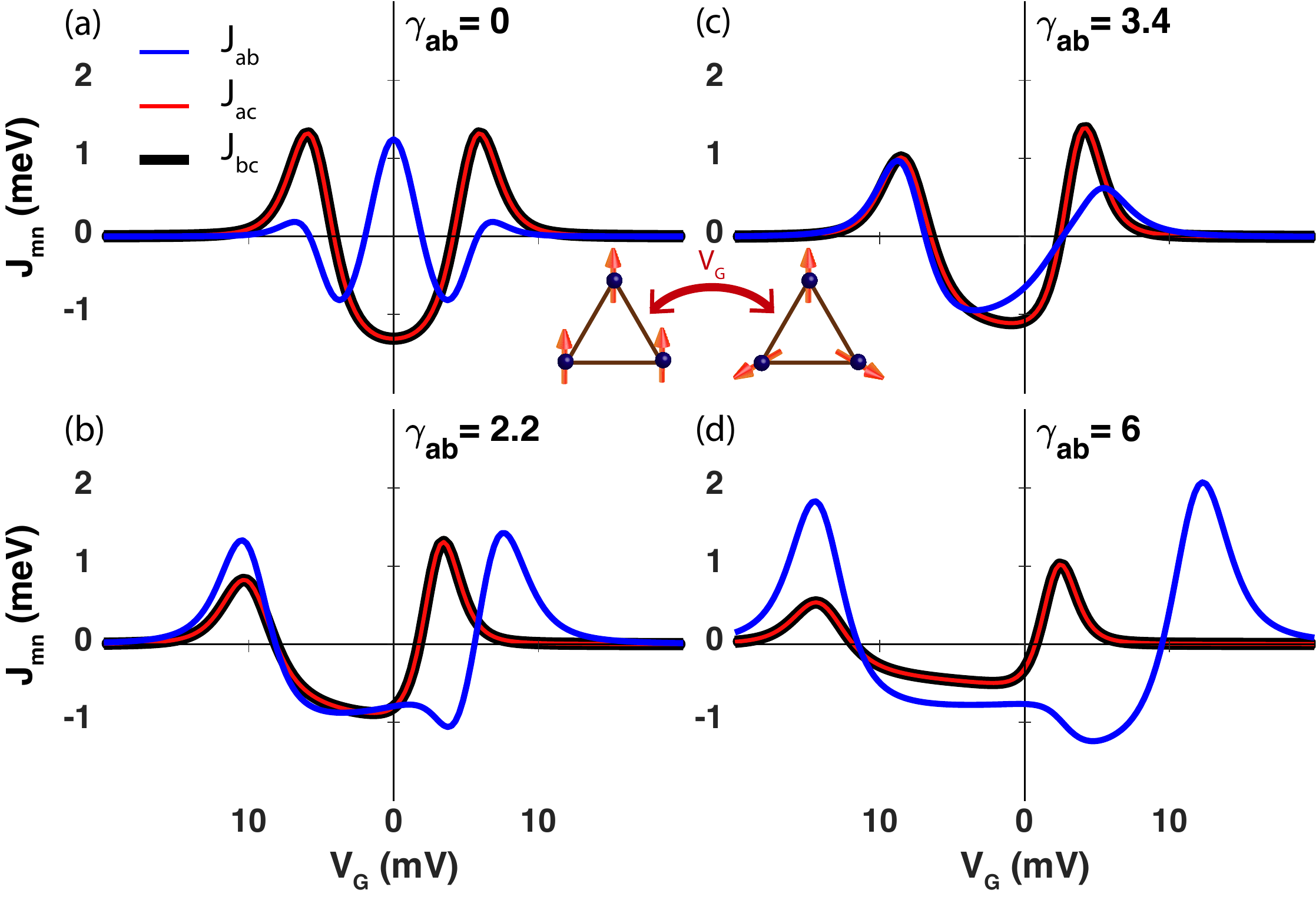}
	\caption{Switching dynamics among ordered spin states as function of the applied gate votlage $V_\text{G}$ at zero voltage bias.}
	\label{frust3}
\end{figure}

By staring at the contour plots in Fig. \ref{frust6} it can be seen that there are regions in the $V_\text{SD}$--$V_\text{G}$ diagram for the effective exchange that correspond to one type of ordering and by the selecting biasing of the junction, a different state can be engineered, following the line of thought of the experiments reported in \cite{Loth2012,Heinrich2013,Heinrich2015, Khajetoorians2011,Khajetoorians2012,Jungwirth2016,Otte2009}, where magnetic excitations in ad-atoms are engineered by the action of electric drives, in many cases in agreement with the RKKY limit. Here we have explored an additional degree of freedom for controlling this tailoring at the atomic scale, that is, the gate field $V_\text{G}$ which provides reasonable tuning possibility among anti-ferromagnetic (AFM) and ferromagnetic (FM) ones as shown in Fig. \ref{frust3}, hence empowering the optimal location of the operating point of the molecule in the magnetic $V_\text{SD}$--$V_\text{G}$ diagram. This means of control, as stated before, has been successfully demonstrated in \cite{Koole2015}. Now, we will focus on what are the possibilities for tuning either anti-ferromagneticaly or ferromagnetically the magnetic units in the probed molecule. Fig. \ref{frust3} considers the effective exchanges among spins as a function of gate voltage for a zero-bias condition. Therefore, this prediction cannot be verified by electrical means, though it provides some insight on how to tune the system with the gate field to commute among magnetic configurations, as exemplified in Fig. \ref{schemes1}.

By looking now at Fig. \ref{frust3}, one sees the gating conditions for which the anti-ferromagnetic state will be more stable and robust against variations in the gate field and against modulations of the parameter determining the degree of quantum interference $\gamma_{ab}$, for which we determined this condition to be $V_\text{G}\approx -4.0$ mV. By changing the gating condition is possible to obtain the all-ferromagnetic configuration for small values of $\gamma_{ab}$, or this can be achieved by stabilizing the gate field around the set value, and rather modulating $\gamma_{ab}$ as shown in the lower right panel of Fig. \ref{frust3}; condition perfectly exemplified in the diagram shown in Fig. \ref{schemes1}.

\section{EigenValue and EigenState Analysis}

In the present paper, we are interested in two regimes of ordering: Ferromagnetic and Anti-Ferromagnetic. To determine whether the spin states exhibit classical correlations or quantum entanglement, we analyze the eigen-values and the corresponding eigen-state and the associated Von-Neumann entropy of the effective spin Hamiltonian \ref{eff}. Let's consider two relevant cases , where $\bm{\mathcal{J}}_{ab}=\bm{\mathcal{J}}_{ac}=\bm{\mathcal{J}}_{bc}=\bm{\mathcal{J}}$ and where $\bm{\mathcal{J}}_{ab}=\bm{\mathcal{J}}$, $\bm{\mathcal{J}}_{ac}=\bm{\mathcal{J}}_{bc}=\bm{\mathcal{J}}_{0}$, and determine the spectral diagram for the quantum spin states in both, the ferromagnetic and the anti-ferromagnetic spin ordering.

\subsection{Ferromagnetic Ordering}

\subsubsection{Case: Equal Spin-Spin Effective Couplings}
From expression \ref{eigen}, the eigen-energies of the Spin Hamiltonian give:
\begin{align}
\mathcal{E}_{spin}^{(1)}&=\mathcal{E}_{spin}^{(2)}=\mathcal{E}_{spin}^{(3)}=\mathcal{E}_{spin}^{(4)}=-\bm{\mathcal{J}};
\label{eigen1}
\\
\mathcal{E}_{spin}^{(5)}&=\mathcal{E}_{spin}^{(6)}=3\bm{\mathcal{J}};
\label{eigen2}
\\
\mathcal{E}_{spin}^{(7)}&=\mathcal{E}_{spin}^{(8)}=3\bm{\mathcal{J}};
\label{eigen3}
\end{align}
where the ground state of the Hamiltonian under this conditions is a quartet state $\bm{\chi}_{q}$ given by:
\begin{align}
\bm{\chi}_{q}=\left[
\begin{array}{c}
\ket{\uparrow\uparrow\uparrow}
\\
\\
\frac{1}{\sqrt{3}}\left(\ket{\uparrow\uparrow\downarrow}+\ket{\uparrow\downarrow\downarrow}+\ket{\downarrow\uparrow\uparrow}\right)
\\
\\
\frac{1}{\sqrt{3}}\left(\ket{\uparrow\downarrow\downarrow}+\ket{\downarrow\uparrow\downarrow}+\ket{\downarrow\downarrow\uparrow}\right)
\\
\\
\ket{\downarrow\downarrow\downarrow}
\end{array}
 \right]
\label{eigen4}
\end{align}
\subsection{Case: Symmetric Coupling I}
For this case, which is when $\bm{\mathcal{J}}_{ab}=\bm{\mathcal{J}}$, $\bm{\mathcal{J}}_{ac}=\bm{\mathcal{J}}_{bc}=\bm{\mathcal{J}}_{0}$, the eigen energies give:
\begin{align}
\mathcal{E}_{spin}^{(1)}=\mathcal{E}_{spin}^{(2)}=\mathcal{E}_{spin}^{(3)}=\mathcal{E}_{spin}^{(4)}&=-\left(\bm{\mathcal{J}}+2\bm{\mathcal{J}}_{0}\right);
\label{eigen5}
\\
\mathcal{E}_{spin}^{(5)}=\mathcal{E}_{spin}^{(6)}&=-\bm{\mathcal{J}}+4\bm{\mathcal{J}}_{0};
\label{eigen6}
\\
\mathcal{E}_{spin}^{(7)}=\mathcal{E}_{spin}^{(8)}&=3\bm{\mathcal{J}};
\label{eigen7}
\end{align}
where the ground state for this case is as well given by \ref{eigen4}.
\subsection{Anti-Ferromagnetic Ordering}
\subsubsection{Case: Equal Spin-Spin Effective Couplings}
\begin{align}
\mathcal{E}_{spin}^{(1)}=\mathcal{E}_{spin}^{(2)}=\mathcal{E}_{spin}^{(3)}=\mathcal{E}_{spin}^{(4)}&=3\bm{\mathcal{J}};
\label{eigen8}
\\
\mathcal{E}_{spin}^{(5)}=\mathcal{E}_{spin}^{(6)}&=-3\bm{\mathcal{J}};
\label{eigen9}
\\
\mathcal{E}_{spin}^{(7)}=\mathcal{E}_{spin}^{(8)}&=-3\bm{\mathcal{J}};
\label{eigen10}
\end{align}
with an associated degenerate doublet as a ground state given by:
\begin{align}
\bm{\chi}_{q_{1}}=\left[
\begin{array}{c}
\frac{1}{\sqrt{1+\gamma^{2}_{\uparrow\uparrow\downarrow}+\gamma^{2}_{\uparrow\downarrow\uparrow}}}
\left(\gamma_{\uparrow\uparrow\downarrow}\ket{\uparrow\uparrow\downarrow}+\gamma_{\uparrow\downarrow\uparrow}\ket{\uparrow\downarrow\uparrow}+\ket{\downarrow\uparrow\uparrow}\right)
\\
\\
\frac{1}{\sqrt{1+\gamma^{2}_{\uparrow\downarrow\downarrow}+\gamma^{2}_{\downarrow\uparrow\downarrow}}}
\left(\gamma_{\uparrow\downarrow\downarrow}\ket{\uparrow\downarrow\downarrow}+\gamma_{\downarrow\uparrow\downarrow}\ket{\downarrow\uparrow\downarrow}+\ket{\downarrow\downarrow\uparrow}\right)
\end{array}
 \right]
\label{eigen15}
\end{align};
\begin{align}
\bm{\chi}_{q_{2}}=\left[
\begin{array}{c}
\frac{1}{\sqrt{1+\gamma^{2}_{\uparrow\uparrow\downarrow}+\gamma^{2}_{\uparrow\downarrow\uparrow}}}
\left(\gamma_{\uparrow\uparrow\downarrow}\ket{\uparrow\uparrow\downarrow}+\gamma_{\uparrow\downarrow\uparrow}\ket{\uparrow\downarrow\uparrow}+\ket{\downarrow\uparrow\uparrow}\right)
\\
\\
\frac{1}{\sqrt{1+\gamma^{2}_{\uparrow\downarrow\downarrow}+\gamma^{2}_{\downarrow\uparrow\uparrow}}}
\left(\gamma_{\uparrow\downarrow\downarrow}\ket{\uparrow\downarrow\downarrow}+\gamma_{\downarrow\uparrow\uparrow}\ket{\downarrow\uparrow\uparrow}+\ket{\downarrow\downarrow\uparrow}\right)
\end{array}
 \right]
\label{eigen11b}
\end{align}
\subsection{Case: Symmetric Coupling}
\begin{align}
\mathcal{E}_{spin}^{(1)}=\mathcal{E}_{spin}^{(2)}=\mathcal{E}_{spin}^{(3)}=\mathcal{E}_{spin}^{(4)}&=\bm{\mathcal{J}}+2\bm{\mathcal{J}}_{0};
\label{eigen12}
\\
\mathcal{E}_{spin}^{(5)}=\mathcal{E}_{spin}^{(6)}&=-3\bm{\mathcal{J}};
\label{eigen13}
\\
\mathcal{E}_{spin}^{(7)}=\mathcal{E}_{spin}^{(8)}&=\bm{\mathcal{J}}-4\bm{\mathcal{J}}_{0};
\label{eigen14}
\end{align}
The ground state will depend on whether $\bm{\mathcal{J}}>\bm{\mathcal{J}}_{0}$ or $\bm{\mathcal{J}}<\bm{\mathcal{J}}_{0}$. For $\bm{\mathcal{J}}>\bm{\mathcal{J}}_{0}$, the ground state energy is $E_{G.S}=-3\bm{\mathcal{J}}$ and the associated eigen-state is given by:
\begin{align}
\bm{\chi}_{q}=\left[
\begin{array}{c}
\frac{1}{\sqrt{1+\gamma^{2}_{\uparrow\uparrow\downarrow}+\gamma^{2}_{\uparrow\downarrow\uparrow}}}
\left(\gamma_{\uparrow\uparrow\downarrow}\ket{\uparrow\uparrow\downarrow}+\gamma_{\uparrow\downarrow\uparrow}\ket{\uparrow\downarrow\uparrow}+\ket{\downarrow\uparrow\uparrow}\right)
\\
\\
\frac{1}{\sqrt{1+\gamma^{2}_{\uparrow\downarrow\downarrow}+\gamma^{2}_{\downarrow\uparrow\downarrow}}}
\left(\gamma_{\uparrow\downarrow\downarrow}\ket{\uparrow\downarrow\downarrow}+\gamma_{\downarrow\uparrow\downarrow}\ket{\downarrow\uparrow\downarrow}+\ket{\downarrow\downarrow\uparrow}\right)
\end{array}
 \right]
\label{eigen152}
\end{align}
For $\bm{\mathcal{J}}<\bm{\mathcal{J}}_{0}$, the ground state energy is $E_{G.S}=\bm{\mathcal{J}}-4\bm{\mathcal{J}}_{0}$ and the associated eigen-state is given by
\begin{align}
\bm{\chi}_{q}=\left[
\begin{array}{c}
\frac{1}{\sqrt{1+\gamma^{2}_{\uparrow\uparrow\downarrow}+\gamma^{2}_{\uparrow\downarrow\uparrow}}}
\left(\gamma_{\uparrow\uparrow\downarrow}\ket{\uparrow\uparrow\downarrow}+\gamma_{\uparrow\downarrow\uparrow}\ket{\uparrow\downarrow\uparrow}+\ket{\downarrow\uparrow\uparrow}\right)
\\
\\
\frac{1}{\sqrt{1+\gamma^{2}_{\uparrow\downarrow\downarrow}+\gamma^{2}_{\downarrow\uparrow\uparrow}}}
\left(\gamma_{\uparrow\downarrow\downarrow}\ket{\uparrow\downarrow\downarrow}+\gamma_{\downarrow\uparrow\uparrow}\ket{\downarrow\uparrow\uparrow}+\ket{\downarrow\downarrow\uparrow}\right)
\end{array}
\right]
\label{eigen16}
\end{align}
\section{Conclusions}
In the present reported work we considered a magnetic trimer, with a three level electronic system coupled with its hosted magnetic units via Kondo interaction, driven by a metallic tunneling junction, such that resembles a scanning tunneling microscopy experiment on magnetic ad-atoms on metallic surface. Additionally, the set up allows for an electric field acting as a gate drive, which in convergence with the bias voltage $V_{DS}$ define a magnetic $V_{DS}$--$V_{G}$ diagram where the symmetry status of the magnetic trimer becomes evident. We have shown that through the following three different mechanisms:
\begin{enumerate}
	\item Modulation of the nature and strength of the electronic quantum interference,
	\item Voltage bias induced non-equilibrium stationary dynamics,
	\item Gate field control of the electronic structure \cite{Koole2015}
\end{enumerate}
a switching dynamics between all anti-ferromagnetic coupling spin state and all ferromagnetic one can be induced. Interestingly, the results reported for the indirect exchange interaction shown in Figs. \ref{model6},\ref{frust6},\ref{frust3}, showed that not only the orientation but the relative strength or the order can be tuned in the magnetic $V_{DS}-V_{G}$ diagram for a variety of modulations of the parameter $\gamma_{ab}$, showing the corresponding of the magnetic formation with respect to the control means proposed in this work. Lastly, focusing on the objective of the study, which is to trace a correlation between ordering in the magnetic molecule and coherence in the electronic background, we infer from Fig. \ref{ttx} that around the set Fermi level by the Gate field $V_{G}$ (dotted line in black), there is a controlled but prominent (Fano-like dip?) decay in the electronic transmission signing the emergence of the quantum interference of destructive nature. This conclusion, suggests that in the ferromagnetic induced delocalization, the ability to exhibit quantum interference dominates over the quantum coherent phenomena induced by the anti-ferromagnetic localization, in which case entropy driven processes will tend to be robust against anti-ferromagnetic induced electronic decoherence. This conclusion is in agreement with the predictions published in \cite{Jaramillo2017} with regards to the competition between singlet-triplet formation and orbital localization.
\section{Acknowledgments}
We would like to acknowledge M. Araujo, L. Nordstr\"om, P. Oppeneer, M. Pereiro, and Y. Sassa for useful discussions and feedback. We acknowledge funding from Minciencias (Ministry of Science, Technology and Innovation - Colombia) and From Universidad del Valle through the grant for the project with reference CI 71261 corresponding to the call by Minciencias No. 649 from 2019 for postdoctoral research scholars under the agreement number 80740-618-2020.


\appendix
\begin{widetext}
\section{Evaluation of the Retarded Green's Function for the Molecular Trimer}
\label{green_func}
The Green's function, in its retarded form, can be derived from the equation of motion \cite{Jauho1994,JauhoKinetics} technique as defined for the Keldysh contour, yielding:
\begin{align}
G_{mn\sigma\sigma'}^{R}(t,t')=\delta_{mn}\delta_{\sigma\sigma'}\mathcal{G}^{R}_{m\sigma}(t,t')&
+\sum_{m_{1}}\gamma_{mm_{1}}\int\mathcal{G}^{R}_{m\sigma}(t,\tau)G^{R}_{m_{1}n\sigma\sigma'}(\tau,t')d\tau
+\sum_{\sigma_{1}}J_{m}\int\bm{\sigma}_{\sigma\sigma_{1}}\cdot\left\langle  \bm{S}_{m}(\tau)\right\rangle
\mathcal{G}^{R}_{m\sigma}(t,\tau)G^{R}_{mn\sigma_{1}\sigma'}(\tau,t')d\tau
\nonumber
\\
&
+\sum_{\chi\mu}\int \int\mathcal{G}^{R}_{m\sigma}(t,\tau)\Sigma^{R(\chi)}_{m\mu \sigma\sigma}(\tau,\tau')G^{R}_{\mu n\sigma\sigma'}(\tau',t')d\tau'd\tau,
\label{alla2}
\end{align}
where $\mathcal{G}^{R}_{m\sigma}(t,t')$ satisfying the Schr\"odinger like equation given by:
$$
\left(i\hbar\frac{\partial}{\partial t}-\bar{\epsilon}_{m\sigma}\right)\mathcal{G}_{m\sigma}(t,t')=\delta(t-t').
$$
From the model described in Eqs. (\ref{ham}), (\ref{alla2}) defines $\left[G^{R}_{mn\sigma\sigma'}(\omega)\right]^{-1}$ such that $G^{R}_{mn\sigma\sigma'}(\omega)\left[G^{R}_{mn\sigma\sigma'}(\omega)\right]^{-1}=\delta_{mn}\delta_{\sigma\sigma'}$. Thereafter, $\left[G^{R}_{mn\sigma\sigma'}(\omega)\right]^{-1}$ can be written in matrix form in the following way:
\begin{align}
\left[G^{R}_{mn\sigma\sigma'}(\omega)\right]^{-1}=
\left[
\begin{array}{ccc}
\hbar\omega-\bar{\epsilon}_{a\sigma}+\frac{i}{2}\Gamma_{a\sigma}&-\gamma_{ab}&-\left(\gamma_{ac}-\frac{i}{2}\Gamma_{ac\sigma}\right)
\\
\\
-\gamma_{ba}&\hbar\omega-\bar{\epsilon}_{b\sigma}+\frac{i}{2}\Gamma_{b\sigma}&-\left(\gamma_{bc}-\frac{i}{2}\Gamma_{bc\sigma}\right)
\\
\\
-\left(\gamma_{ca}-\frac{i}{2}\Gamma_{ca\sigma}\right)&-\left(\gamma_{cb}-\frac{i}{2}\Gamma_{cb\sigma}\right)&\hbar\omega-\bar{\epsilon}_{c\sigma}+\frac{i}{2}\Gamma_{c\sigma}
\end{array}
\right],
\label{spin3}
\end{align}
or in Dyson equation form (self-energies $\bm{\Gamma}^{(\alpha)}_{\sigma\sigma}$ and $\bm{\Gamma}^{(\beta)}_{\sigma\sigma}$ become evident):
\begin{equation}
\bm{G}^{R}(\omega)=\left(\bm{\Omega}-[\bm{\gamma}]+\frac{i}{\hbar}\left(\bm{\Gamma}^{(\alpha)}_{\sigma\sigma}+\bm{\Gamma}^{(\beta)}_{\sigma\sigma}\right)\right)^{-1},
\label{inv}
\end{equation}
Where the matrices $\bm{\Omega}$, $[\bm{\gamma}]$, $\bm{\Gamma}^{(\alpha)}_{\sigma\sigma}$ and $\bm{\Gamma}^{(\beta)}_{\sigma\sigma}$ are given by:
\begin{equation}
\begin{array}{cccc}
\hspace{-2.5cm}
\bm{\Omega}=\left[
\begin{array}{ccc}
\bar{\epsilon}_{a\sigma}&0&0\\
\\
0&\bar{\epsilon}_{b\sigma}&0\\
\\
0&0&\bar{\epsilon}_{c\sigma}
\end{array}\right],
&
\left[\bm{\gamma}\right]=\left[
\begin{array}{ccc}
0&\gamma_{ab}&\gamma_{ac}\\
\\
\gamma_{ba}&0&\gamma_{bc}\\
\\
\gamma_{ca}&\gamma_{cb}&0
\end{array}\right],
&
\bm{\Gamma}^{(\alpha)}_{\sigma\sigma}=\left[
\begin{array}{ccc}
\Gamma^{(\alpha)}_{a\sigma}&0&\Gamma^{(\alpha)}_{ac\sigma}\\
\\
0&0&0\\
\\
\Gamma^{(\alpha)}_{ca\sigma}&0&\Gamma^{(\alpha)}_{c\sigma}
\end{array}\right],
&
\bm{\Gamma}^{(\beta)}_{\sigma\sigma}=\left[
\begin{array}{ccc}
0&0&0\\
\\
0&\Gamma^{(\beta)}_{b\sigma}&\Gamma^{(\beta)}_{bc\sigma}\\
\\
\\
0&\Gamma^{(\beta)}_{cb\sigma}&\Gamma^{(\beta)}_{c\sigma}
\end{array}\right],
\end{array}
\end{equation}
and the renormalized energies in matrix $\bm{\Omega}$ are defined according to:
\begin{equation}
\bar{\epsilon}_{m\sigma}=\epsilon_{m\sigma}+J_{m} \bm{\sigma}^{(z)}_{\sigma\sigma}\left\langle S_{m}\right\rangle.
\end{equation}
\section{Effective Spin-Spin Hamiltonian}
The effective spin $\frac{1}{2}$ Hamiltonian for a spin trimer is written in the following way:
\begin{equation}
\bm{\mathcal{H}}_{spin}=\mathcal{J}_{ab}\bm{S}_{a}\cdot\bm{S}_{b}+\mathcal{J}_{ac}\bm{S}_{a}\cdot\bm{S}_{c}+\mathcal{J}_{bc}\bm{S}_{b}\cdot\bm{S}_{c},
\label{eff} 
\end{equation}
where $\mathcal{J}_{mn}$ is given according to \cite{Fransson2014} as:
\begin{equation}
\mathcal{J}_{mn}=\frac{J_{m}J_{n}}{2}\int\int\frac{G^{(0)<}_{mn}(\epsilon)G^{(0)>}_{nm}(\epsilon')-G^{(0)>}_{mn}(\epsilon)G^{(0)<}_{nm}(\epsilon')-\bm{G}^{(1)<}_{mn}(\epsilon)\cdot\bm{G}^{(1)>}_{nm}(\epsilon')+\bm{G}^{(1)>}_{mn}(\epsilon)\cdot\bm{G}^{(1)<}_{nm}(\epsilon')}{\hbar\omega-\epsilon+\epsilon'}
\frac{d\epsilon}{2\pi}\frac{d\epsilon'}{2\pi}.
\label{exchange}
\end{equation}
The spin dot products shown in Eq. (\ref{eff}), can be expanded as a complete Hilbert space according to the following tensor products:
\begin{align}
\bm{S}_{a}\cdot\bm{S}_{b}&=S_{ax}\otimes S_{bx}\otimes \mathbb{I}_{2\times 2}+S_{ay}\otimes S_{by}\otimes \mathbb{I}_{2\times 2}+S_{az}\otimes S_{bz}\otimes \mathbb{I}_{2\times 2},
\label{tensor1}
\\
\bm{S}_{b}\cdot\bm{S}_{c}&=\mathbb{I}_{2\times 2}\otimes S_{bx}\otimes S_{cx}+\mathbb{I}_{2\times 2}\otimes S_{by}\otimes S_{cy}+\mathbb{I}_{2\times 2}\otimes S_{bz}\otimes S_{cz},
\label{tensor2}
\\
\bm{S}_{a}\cdot\bm{S}_{c}&=S_{ax}\otimes \mathbb{I}_{2\times 2}\otimes S_{cx}+S_{ay}\otimes \mathbb{I}_{2\times 2} \otimes S_{cy}+S_{az}\otimes \mathbb{I}_{2\times 2}\otimes S_{cz},
\label{tensor3}
\end{align}
where the operators $S_{ix}$, $S_{iy}$ and $S_{iz}$ for a spin $\frac{1}{2}$ degree of freedom are given the well known $\sigma$ matrices.
\begin{equation}
\bm{\mathcal{H}}_{spin}=
\left[\begin{array}{cccccccc}
\mathcal{J}_{+++}&0&0&0&0&0&0&0\\
0&\mathcal{J}_{+--}&2\mathcal{J}_{bc}&0&2\mathcal{J}_{ac}&0&0&0\\
0&2\mathcal{J}_{bc}&\mathcal{J}_{-+-}&0&2\mathcal{J}_{ab}&0&0&0\\
0&0&0&\mathcal{J}_{--+}&0&2\mathcal{J}_{ab}&2\mathcal{J}_{ac}&0\\
0&2\mathcal{J}_{ac}&2\mathcal{J}_{ab}&0&\mathcal{J}_{--+}&0&0&0\\
0&0&0&2\mathcal{J}_{ab}&0&\mathcal{J}_{-+-}&2\mathcal{J}_{bc}&0\\
0&0&0&2\mathcal{J}_{ac}&0&2\mathcal{J}_{bc}&\mathcal{J}_{+--}&0\\
0&0&0&0&0&0&0&\mathcal{J}_{+++}.\\
\end{array}
\right]
\label{eff1}
\end{equation}
The diagonal elements of the above Eq. are given by:
\begin{align}
\mathcal{J}_{+++}&=\mathcal{J}_{ab}+\mathcal{J}_{ac}+\mathcal{J}_{bc},
\nonumber
\\
\mathcal{J}_{+--}&=\mathcal{J}_{ab}-\mathcal{J}_{ac}-\mathcal{J}_{bc},
\nonumber
\\
\mathcal{J}_{--+}&=-\mathcal{J}_{ab}-\mathcal{J}_{ac}+\mathcal{J}_{bc},
\nonumber
\\
\mathcal{J}_{-+-}&=-\mathcal{J}_{ab}+\mathcal{J}_{ac}-\mathcal{J}_{bc}.
\nonumber
\end{align}
Eq. \ref{eff1}, can be written as a block matrix of $2\times 2$, $1\times 1$, $1\times 3$ (zeros) and $3\times 1$ (zeros) matrices as elements:
\begin{equation}
\bm{\mathcal{H}}_{spin}=
\left[\begin{array}{cccc}
\mathcal{J}_{11}&\bm{0}_{1\times 3}&\bm{0}_{1\times 3}&0\\
\bm{0}_{3\times 1}&\mathcal{J}_{22}&\mathcal{J}_{23}&\bm{0}_{3\times 1}\\
\bm{0}_{3\times 1}&\mathcal{J}_{32}&\mathcal{J}_{33}&\bm{0}_{3\times 1}\\
0&\bm{0}_{1\times 3}&\bm{0}_{1\times 3}&\mathcal{J}_{44}\\
\end{array}
\right]
\label{eff1a}
\end{equation}
The eigen-value problem, $\left|\bm{\mathcal{H}}_{spin}-\lambda\mathbb{I}_{8\times 8}\right|=0$, proceeds as follows:
\begin{align}
\left|\bm{\mathcal{H}}_{spin}-\lambda\mathbb{I}_{8\times 8}\right|&=
\left|\left[\begin{array}{cccc}
\mathcal{J}_{11}&\bm{0}_{1\times 3}&\bm{0}_{1\times 3}&0\\
\bm{0}_{3\times 1}&\mathcal{J}_{22}&\mathcal{J}_{23}&\bm{0}_{3\times 1}\\
\bm{0}_{3\times 1}&\mathcal{J}_{32}&\mathcal{J}_{33}&\bm{0}_{3\times 1}\\
0&\bm{0}_{1\times 3}&\bm{0}_{1\times 3}&\mathcal{J}_{44}
\end{array}\right]
-
\left[\begin{array}{cccc}
\lambda&\bm{0}_{1\times 3}&\bm{0}_{1\times 3}&0\\
\bm{0}_{3\times 1}&\lambda\mathbb{I}_{3\times 3}&\bm{0}_{3\times 3}&\bm{0}_{3\times 1}\\
\bm{0}_{3\times 1}&\bm{0}_{3\times 3}&\lambda\mathbb{I}_{3\times 3}&\bm{0}_{3\times 1}\\
0&\bm{0}_{1\times 3}&\bm{0}_{1\times 3}&\lambda
\end{array}\right]\right|=0,
\nonumber
\\
&=\left|\left[\begin{array}{cccc}
\mathcal{J}_{11}-\lambda&\bm{0}_{1\times 3}&\bm{0}_{1\times 3}&0\\
\bm{0}_{3\times 1}&\mathcal{J}_{22}-\lambda\mathbb{I}_{3\times 3}&\mathcal{J}_{23}&\bm{0}_{3\times 1}\\
\bm{0}_{3\times 1}&\mathcal{J}_{32}&\mathcal{J}_{33}-\lambda\mathbb{I}_{3\times 3}&\bm{0}_{3\times 1}\\
0&\bm{0}_{1\times 3}&\bm{0}_{1\times 3}&\mathcal{J}_{44}-\lambda
\end{array}\right]\right|=0,
\nonumber
\\
&=\left(\mathcal{J}_{11}-\lambda\right)
\left|\left[\begin{array}{ccc}
\mathcal{J}_{22}-\lambda\mathbb{I}_{3\times 3}&\mathcal{J}_{23}&\bm{0}_{3\times 1}\\
\mathcal{J}_{32}&\mathcal{J}_{33}-\lambda\mathbb{I}_{3\times 3}&\bm{0}_{3\times 1}\\
\bm{0}_{1\times 3}&\bm{0}_{1\times 3}&\mathcal{J}_{44}-\lambda
\end{array}\right]\right|=0,
\nonumber
\\
&=\left(\mathcal{J}_{11}-\lambda\right)\left|\left[\left(\mathcal{J}_{22}-\lambda\mathbb{I}_{3\times 3}\right)\left(\mathcal{J}_{33}-\lambda\mathbb{I}_{3\times 3}\right)\left(\mathcal{J}_{44}-\lambda\right)-\mathcal{J}_{23}\mathcal{J}_{32}\left(\mathcal{J}_{44}-\lambda\right)\right]\right|=0,
\nonumber
\\
&=\left(\mathcal{J}_{11}-\lambda\right)\left(\mathcal{J}_{44}-\lambda\right)\left|\left[\left(\mathcal{J}_{22}-\lambda\mathbb{I}_{3\times 3}\right)\left(\mathcal{J}_{33}-\lambda\mathbb{I}_{3\times 3}\right)-\mathcal{J}_{23}\mathcal{J}_{32}\right]\right|=0.
\label{eigen1A}
\end{align}
From Eq. \ref{eff1}, matrices $\mathcal{J}_{22}$, $\mathcal{J}_{33}$, $\mathcal{J}_{23}$ and $\mathcal{J}_{32}$ are given by:
\begin{align}
\hspace{-0.5cm}
\begin{array}{cccc}
\mathcal{J}_{22}=
\left[\begin{array}{ccc}
\mathcal{J}_{+--}&2\mathcal{J}_{bc}&0\\
2\mathcal{J}_{bc}&\mathcal{J}_{-+-}&0\\
0&0&\mathcal{J}_{--+}\\
\end{array}
\right],&
\mathcal{J}_{33}=
\left[\begin{array}{ccc}
\mathcal{J}_{--+}&0&0\\
0&\mathcal{J}_{-+-}&2\mathcal{J}_{bc}\\
0&2\mathcal{J}_{bc}&\mathcal{J}_{+--}\\
\end{array}
\right],&
\mathcal{J}_{23}=
\left[\begin{array}{ccc}
2\mathcal{J}_{ac}&0&0\\
2\mathcal{J}_{ab}&0&0\\
0&2\mathcal{J}_{ab}&2\mathcal{J}_{ac}\\
\end{array}
\right],
&
\mathcal{J}_{32}=
\left[\begin{array}{ccc}
2\mathcal{J}_{ac}&2\mathcal{J}_{ab}&0\\
0&0&2\mathcal{J}_{ab}\\
0&0&2\mathcal{J}_{ac}\\
\end{array}
\right].
\end{array}
\end{align}
The term $\mathcal{J}_{23}\mathcal{J}_{32}$ in Eq. \ref{eigen1A} can be further elaborated as follows:
\begin{align}
\mathcal{J}_{23}\mathcal{J}_{32}&=
\left[\begin{array}{ccc}
2\mathcal{J}_{ac}&0&0\\
2\mathcal{J}_{ab}&0&0\\
0&2\mathcal{J}_{ab}&2\mathcal{J}_{ac}\\
\end{array}
\right]
\left[\begin{array}{ccc}
2\mathcal{J}_{ac}&2\mathcal{J}_{ab}&0\\
0&0&2\mathcal{J}_{ab}\\
0&0&2\mathcal{J}_{ac}\\
\end{array}
\right]=
\left[\begin{array}{ccc}
4\mathcal{J}^{2}_{ac}&4\mathcal{J}_{ac}\mathcal{J}_{ab}&0\\
4\mathcal{J}_{ac}\mathcal{J}_{ab}&4\mathcal{J}^{2}_{ab}&0\\
0&0&4\left(\mathcal{J}^{2}_{ac}+\mathcal{J}^{2}_{ab}\right),\\
\end{array}
\right]
\end{align} 
and the term $\left(\mathcal{J}_{22}-\lambda\mathbb{I}_{3\times 3}\right)\left(\mathcal{J}_{33}-\lambda\mathbb{I}_{3\times 3}\right)$ is given by:
\begin{align}
\left(\mathcal{J}_{22}-\lambda\mathbb{I}_{3\times 3}\right)\left(\mathcal{J}_{33}-\lambda\mathbb{I}_{3\times 3}\right)&=
\left[\begin{array}{ccc}
\mathcal{J}_{+--}-\lambda&2\mathcal{J}_{bc}&0\\
2\mathcal{J}_{bc}&\mathcal{J}_{-+-}-\lambda&0\\
0&0&\mathcal{J}_{--+}-\lambda\\
\end{array}
\right]
\left[\begin{array}{ccc}
\mathcal{J}_{--+}-\lambda&0&0\\
0&\mathcal{J}_{-+-}-\lambda&2\mathcal{J}_{bc}\\
0&2\mathcal{J}_{bc}&\mathcal{J}_{+--}-\lambda\\
\end{array}
\right],
\nonumber
\\
&=\left[\begin{array}{ccc}
\left(\mathcal{J}_{+--}-\lambda\right)\left(\mathcal{J}_{--+}-\lambda\right)&2\mathcal{J}_{bc}\left(\mathcal{J}_{-+-}-\lambda\right)&4\mathcal{J}^{2}_{bc}\\
2\mathcal{J}_{bc}\left(\mathcal{J}_{--+}-\lambda\right)&\left(\mathcal{J}_{-+-}-\lambda\right)^{2}&2\mathcal{J}_{bc}\left(\mathcal{J}_{-+-}-\lambda\right)\\
0&2\mathcal{J}_{bc}\left(\mathcal{J}_{--+}-\lambda\right)&\left(\mathcal{J}_{--+}-\lambda\right)\left(\mathcal{J}_{+--}-\lambda\right)\\
\end{array}
\right],
\end{align}
and replacing the latter and former result in Eq. \ref{eigen1A}, the eigen-value problem now reads:
\begin{align}
\left|\bm{\mathcal{H}}_{spin}-\lambda\mathbb{I}_{8\times 8}\right|&=\left(\mathcal{J}_{11}-\lambda\right)\left(\mathcal{J}_{44}-\lambda\right)\left|\left[\left(\mathcal{J}_{22}-\lambda\mathbb{I}_{3\times 3}\right)\left(\mathcal{J}_{33}-\lambda\mathbb{I}_{3\times 3}\right)-\mathcal{J}_{23}\mathcal{J}_{32}\right]\right|=0,
\nonumber
\\
&=\left(\mathcal{J}_{11}-\lambda\right)\left(\mathcal{J}_{44}-\lambda\right)
\nonumber
\\
&\times
\left|\left[\begin{array}{ccc}
\left(\mathcal{J}_{+--}-\lambda\right)\left(\mathcal{J}_{--+}-\lambda\right)-4\mathcal{J}^{2}_{ac}&2\mathcal{J}_{bc}\left(\mathcal{J}_{-+-}-\lambda\right)-4\mathcal{J}_{ac}\mathcal{J}_{ab}&4\mathcal{J}^{2}_{bc}\\
\\
2\mathcal{J}_{bc}\left(\mathcal{J}_{--+}-\lambda\right)-4\mathcal{J}_{ac}\mathcal{J}_{ab}&\left(\mathcal{J}_{-+-}-\lambda\right)^{2}-4\mathcal{J}^{2}_{ab}&2\mathcal{J}_{bc}\left(\mathcal{J}_{-+-}-\lambda\right)\\
\\
0&2\mathcal{J}_{bc}\left(\mathcal{J}_{--+}-\lambda\right)&\left(\mathcal{J}_{--+}-\lambda\right)\left(\mathcal{J}_{+--}-\lambda\right)-4\left(\mathcal{J}^{2}_{ac}+\mathcal{J}^{2}_{ab}\right)\\
\end{array}
\right]\right|=0
\end{align}
Eigen-values and associated Eigen-vectors corresponding to the Quartet and two non-degenerate Doublet states are given by:
\begin{align}
\begin{array}{cc}
\mathcal{E}_{spin}^{(1)}=\mathcal{J}_{ab}+\mathcal{J}_{ac}+\mathcal{J}_{bc};&\ket{\phi_{1}}=\ket{\uparrow\uparrow\uparrow}
\\
\\
\mathcal{E}_{spin}^{(2)}=\mathcal{J}_{ab}+\mathcal{J}_{ac}+\mathcal{J}_{bc};
&\ket{\phi_{2}}=\frac{1}{\sqrt{3}}\left(\ket{\uparrow\uparrow\downarrow}+\ket{\uparrow\downarrow\downarrow}+\ket{\downarrow\uparrow\uparrow}\right)
\\
\\
\mathcal{E}_{spin}^{(3)}=\mathcal{J}_{ab}+\mathcal{J}_{ac}+\mathcal{J}_{bc};&
\ket{\phi_{3}}=\frac{1}{\sqrt{3}}\left(\ket{\uparrow\downarrow\downarrow}+\ket{\downarrow\uparrow\downarrow}+\ket{\downarrow\downarrow\uparrow}\right)
\\
\\
\mathcal{E}_{spin}^{(4)}=\mathcal{J}_{ab}+\mathcal{J}_{ac}+\mathcal{J}_{bc};&\ket{\phi_{4}}=\ket{\downarrow\downarrow\downarrow}
\\
\\
\mathcal{E}_{spin}^{(5)}=-\mathcal{J}_{ab}-\mathcal{J}_{ac}-\mathcal{J}_{bc}-2\sqrt{\mathcal{J}_{ab}^{2}+\mathcal{J}_{ac}^{2}+\mathcal{J}_{bc}^{2}-
\mathcal{J}_{ab}\mathcal{J}_{ac}-\mathcal{J}_{ab}\mathcal{J}_{bc}-\mathcal{J}_{bc}\mathcal{J}_{ac}};&
\ket{\phi_{5}}=\frac{\gamma_{\uparrow\uparrow\downarrow}\ket{\uparrow\uparrow\downarrow}+\gamma_{\uparrow\downarrow\uparrow}\ket{\uparrow\downarrow\uparrow}+\ket{\downarrow\uparrow\uparrow}}{\sqrt{1+\gamma^{2}_{\uparrow\uparrow\downarrow}+\gamma^{2}_{\uparrow\downarrow\uparrow}}}
\\
\\
\mathcal{E}_{spin}^{(6)}=-\mathcal{J}_{ab}-\mathcal{J}_{ac}-\mathcal{J}_{bc}-2\sqrt{\mathcal{J}_{ab}^{2}+\mathcal{J}_{ac}^{2}+\mathcal{J}_{bc}^{2}-
\mathcal{J}_{ab}\mathcal{J}_{ac}-\mathcal{J}_{ab}\mathcal{J}_{bc}-\mathcal{J}_{bc}\mathcal{J}_{ac}};&
\ket{\phi_{6}}=\frac{\gamma_{\uparrow\downarrow\downarrow}\ket{\uparrow\downarrow\downarrow}+\gamma_{\downarrow\uparrow\downarrow}\ket{\downarrow\uparrow\downarrow}+\ket{\downarrow\downarrow\uparrow}}{\sqrt{1+\gamma^{2}_{\uparrow\downarrow\downarrow}+\gamma^{2}_{\downarrow\uparrow\downarrow}}}
\\
\\
\mathcal{E}_{spin}^{(7)}=-\mathcal{J}_{ab}-\mathcal{J}_{ac}-\mathcal{J}_{bc}+2\sqrt{\mathcal{J}_{ab}^{2}+\mathcal{J}_{ac}^{2}+\mathcal{J}_{bc}^{2}-
\mathcal{J}_{ab}\mathcal{J}_{ac}-\mathcal{J}_{ab}\mathcal{J}_{bc}-\mathcal{J}_{bc}\mathcal{J}_{ac}};&
\ket{\phi_{7}}=\frac{\gamma_{\uparrow\uparrow\downarrow}\ket{\uparrow\uparrow\downarrow}+\gamma_{\uparrow\downarrow\uparrow}\ket{\uparrow\downarrow\uparrow}+\ket{\downarrow\uparrow\uparrow}}{\sqrt{1+\gamma^{2}_{\uparrow\uparrow\downarrow}+\gamma^{2}_{\uparrow\downarrow\uparrow}}}
\\
\\
\mathcal{E}_{spin}^{(8)}=-\mathcal{J}_{ab}-\mathcal{J}_{ac}-\mathcal{J}_{bc}+2\sqrt{\mathcal{J}_{ab}^{2}+\mathcal{J}_{ac}^{2}+\mathcal{J}_{bc}^{2}-
\mathcal{J}_{ab}\mathcal{J}_{ac}-\mathcal{J}_{ab}\mathcal{J}_{bc}-\mathcal{J}_{bc}\mathcal{J}_{ac}};&
\ket{\phi_{8}}=\frac{\gamma_{\uparrow\downarrow\downarrow}\ket{\uparrow\downarrow\downarrow}+\gamma_{\downarrow\uparrow\uparrow}\ket{\downarrow\uparrow\uparrow}+\ket{\downarrow\downarrow\uparrow}}{\sqrt{1+\gamma^{2}_{\uparrow\downarrow\downarrow}+\gamma^{2}_{\downarrow\uparrow\uparrow}}}
\end{array}
\label{eigen}
\end{align}
Moreover, the aim in this context is to evaluate the spin expectation values of the form $\left\langle \bm{S}_{m}\right\rangle$, for $m=a,b,c$, to then feed them into the retarded Green's function given by Eq. (\ref{inv}), or in the inverse retarded Green's function given by Eq. (\ref{spin3}) in the spirit of the work we presented in \cite{Jaramillo2017,Saygun2016}. To move forward in that department, we employ the definition of the thermal expectation value given by:
\begin{equation}
\label{B14}
\left\langle \bm{S}_{m}\right\rangle=\frac{1}{\mathcal{Z}_{s}}\mathbb{TR}\left[e^{-\beta\bm{\mathcal{H}}_{spin}}\bm{S}_{m\bot}\right],
\end{equation}
where the operator $\bm{S}_{m\bot}$, is the projection of the total spin operator onto the Hilbert space of spin $\bm{S}_{m}$, and $\mathcal{Z}_{s}$ is the partition function of the spin sub-system. Additionally, to fully determine the formation of Quartet and Doublet states for the antiferromagnetic and ferromagnetic ordering case, we calculate the elements of the spin density matrix $\bm{\rho}_{spin}$ in a diagonal basis as follows:
\begin{equation}
\bm{\rho}_{spin}=\frac{e^{-\beta\bm{\mathcal{\bar{H}}}_{spin}}}{\mathcal{Z}_{s}},
\end{equation}
where $\bm{\mathcal{\bar{H}}}_{spin}$ is the Hamiltonian described in Eq. (\ref{eff1}) in diagonal basis.
\section{Differential Charge Conductance}
\label{ConductanceCalculation}
Here, we compute the differential charge conductance. As usual, we start from the derivative:
\begin{equation}
 \bm{\sigma}^{\chi} = \frac{\partial J_{\chi}^{e}}{\partial V} 
 \label{conductance}
\end{equation}

It is useful to recall some important quantities when computing the current observables for quantum transport in stationary regime, regarding the flow of particles from both leads to the molecule and vice-versa. First, from the number operator $\hat{N}$ written from an expansion of the field operator in position and spin basis, we define the current of particles as \cite{Jauho1994}:  
\begin{equation}
 \bm{J}_{\chi}^{N}=\bigg< \frac{d\hat{N}}{dt}  \bigg>
 \label{2}
\end{equation}, where we can see that $\bm{J}_{\chi}^{e}=-e \, \bm{J}_{\chi}^{N}$ can be defined as the current of electrons. Now, from the first law of thermodynamics we write  $\bm{J}_{\chi}^{Q}=\bm{J}_{\chi}^{E} - \mu_{\chi} \bm{J}_{\chi}^{e}$, where $\bm{J}_{\chi}^{Q}$ is the heat current from lead $\chi$ to the molecule, and $\bm{J}_{\chi}^{E}$ is the energy current. The term $\bm{\mu}_{\chi}$ stands for the chemical potential in lead $\chi$ defined from a symmetrical protocol given by $\bm{\mu}_{\chi}=\mu_{o} \pm eV/2$, where $\mu_{o}$ is some reference constant chemical potential and "$V$" is the bias voltage applied to both leads. Analytical expressions are obtained for these currents in Keldish contour using non-equilibrium Green's Functions \cite{Jauho1994}.  For the heat current we have:

\begin{equation}
 \bm{J}_{\chi}^{Q}= - \frac{i}{\hbar} \int \frac{d\epsilon}{2 \pi \hbar}\: (\epsilon - \mu_{\chi}) \: Tr \left\{ \Gamma^{\chi}(\epsilon)\,\left[  G^{>}(\epsilon)f_{\chi}(\epsilon)+G^{<}(\epsilon)f_{\chi}(-\epsilon)   \right] \right\}
 \label{current}
\end{equation} \vspace{0.2cm}

In \eqref{current}, $\bm{\Gamma}_{mn \sigma \sigma}^{\chi}(\epsilon) $ is the Gamma matrix defined in \eqref{self_sigma4} for the couplings with the leads (which is $\bm{\mu}_{\chi}$ independent) and $A(\epsilon)$ is the density of states in the molecule. We note that  $f_{\chi}(-\epsilon)=1-f_{\chi}(\epsilon)$.
We star from the following equations:

\begin{equation}
 \bm{J}_{\chi}^{e} = -ie \int d\epsilon \: Tr \left\{  \Gamma^{\chi}(\epsilon)\,\left[  G^{>}(\epsilon)f_{\chi}(\epsilon)+G^{<}(\epsilon)f_{\chi}(-\epsilon)   \right]  \right\} 
\label{charge} 
\end{equation} \begin{equation}
 G^{>/<}(\epsilon) = G^{R}(\epsilon) \: \Sigma^{>/<}(\epsilon)  \: G^{A}(\epsilon)  
\label{keldish}  
\end{equation} \begin{equation}
 \Sigma^{>/<}(\epsilon) = \mp \: i \sum_{\chi} f_{\chi}(\mp \epsilon) \: \Gamma^{\chi}(\epsilon)
\label{self}
\end{equation}

Eq. \eqref{keldish} is the Keldish equation for $G^{>}$ and $G^{<}$ in energy domain, and \eqref{self} is the self energy of the molecule following the formalism in \cite{Jauho1994}. We obtain an expression for \eqref{conductance} in stationary regime by assuming that both $G^{R, A}$ do not depend on the chemical potential since the problem is not a self-consistent one. An expression for $\partial f_{\chi}(\epsilon)/ \partial \mu_{\chi}=f_{\chi}'(\epsilon)$ is easily obtained as:

 \begin{equation}
    f_{\chi}'(\epsilon) = \frac{\beta}{4} \, Cosh^{-2} \left(  \frac{\beta (\epsilon-\mu_{\chi})}{2}  \right)
    \label{fermideriva}
\end{equation} 

By using \eqref{self} and letting $\Gamma(\epsilon)=\Gamma^{L}(\epsilon)+\Gamma^{L}(\epsilon)$, we write:

\begin{eqnarray}
    \Sigma^{<}(\epsilon) &=&  \: i \sum_{\chi} f_{\chi}(\epsilon) \: \Gamma^{\chi}(\epsilon) =i \left[  \Gamma^{L} f_{L}(\epsilon) +\Gamma^{R} f_{R}(\epsilon)\right] \nonumber \\
    \Sigma^{>}(\epsilon) &=& \Sigma^{<}(\epsilon)-i \,\Gamma(\epsilon) \label{eachselfenergy}
\end{eqnarray} \vspace{0.15cm}

From the symmetric protocol for $\bm{\mu_{\chi}}$ we note that $\frac{\partial \mu_{\chi}}{\partial V} = \pm \frac{e}{2} $ and we can write:

\begin{equation}
    \bm{\sigma}^{\chi} = \frac{\partial \bm{J}_{\chi}^{e}}{\partial \mu_{\chi}} \: \frac{\partial \mu_{\chi}}{\partial V} = \pm \frac{e}{2} \: \frac{\partial \bm{J}_{\chi}^{e}}{\partial \mu_{\chi}} \nonumber 
\end{equation}  
\begin{equation}
    \frac{\partial \bm{J}_{\chi}^{e}}{\partial \mu_{\chi}} = -ie \int d\epsilon \: Tr \left\{  \Gamma^{\chi}(\epsilon)\, \frac{\partial}{\partial \mu_{\chi}}\left[  G^{>}(\epsilon)f_{\chi}(\epsilon)+G^{<}(\epsilon)f_{\chi}(-\epsilon)   \right]  \right\}  
    \label{9}
\end{equation} \vspace{0.2cm}

To elaborate on \eqref{9}, we use Eq. \eqref{keldish} and \eqref{eachselfenergy} to find the derivative:

\begin{equation}
    \frac{\partial G^{>/<}}{\partial \mu_{\chi}}= i\,G^{R}\, \Gamma^{\chi}\,f_{\chi}'\,G^{A}
    \label{10}
\end{equation} 

By using \eqref{10}, we find:

\begin{equation}
  \frac{\partial}{\partial \mu_{\chi}}\left[  G^{>}(\epsilon)f_{\chi}(\epsilon)+G^{<}(\epsilon)f_{\chi}(-\epsilon)  \right]  = f_{\chi}' \Big[ \; G^{>}(\epsilon)-G^{<}(\epsilon)  + i\,G^{R}(\epsilon)\,\Gamma^{\chi}(\epsilon)\,G^{A}(\epsilon) \Big]  
  \label{11}
\end{equation} \vspace{0.2cm}

Finally, we use \eqref{11} in \eqref{9} and plug the result into \eqref{conductance} to obtain:

\begin{equation}
       \bm{\sigma}^{\chi}=\mp \, \frac{i\,e^{2}}{2} \int d\epsilon \:  \frac{\beta}{4} \: \cosh^{-2} \left(  \frac{\beta (\epsilon-\mu_{\chi})}{2}  \right)\:Tr \bigg\{ \;  \Gamma^{\chi}  \Big[ \; G^{>}(\epsilon)-G^{<}(\epsilon) \nonumber + i\,G^{R}(\epsilon)\,\Gamma^{\chi}(\epsilon)\,G^{A}(\epsilon) \Big ]   \bigg\} \nonumber \\ \label{FINAL}
\end{equation}
\pagebreak
\end{widetext}

\bibliography{Bibliography2021}
\end{document}